\documentclass[twocolumn,twocolappendix]{openjournal}

\usepackage{xcolor}
\usepackage{textgreek}
\usepackage[utf8]{inputenc}
\usepackage[english]{babel}
\usepackage{subfig}
\usepackage{amsmath}
\usepackage{natbib}

\usepackage{hyperref}
\hypersetup{
    unicode, 
    colorlinks=true,
    linkcolor=blue,
    citecolor=blue,
    filecolor=blue,
    urlcolor=blue,
}

\usepackage{color,colortbl}
\definecolor{linkcolor}{rgb}{0.0,0.3,0.5}
\usepackage{tensind}
\tensordelimiter{?}
\DeclareGraphicsExtensions{.bmp,.png,.jpg,.pdf}
\usepackage{verbatim}
\usepackage[normalem]{ulem}
\usepackage{orcidlink}
\usepackage{soul}

\newcommand{\bp}[0]{\boldsymbol{p}}

\newcommand{\vecu}{\mathbf{u}}
\newcommand{\vecU}{\mathbf{U}}
\newcommand{\vecE}{\mathbf{E}}
\newcommand{\vecx}{\mathbf{x}}
\newcommand{\vecp}{\mathbf{p}}
\newcommand{\quokka}{{\textsc{Quokka}}}

\newcommand{\edit}[1]{{#1}}

\AtBeginDocument{%

\newcommand{\aref}[1]{\hyperref[#1]{Appendix~\ref{#1}}}
}

\graphicspath{ {./fig/} }

\shorttitle{GPU-accelerated particle-mesh algorithm}
\shortauthors{He, Wibking, Vijayan, Krumholz, \& Li}

\begin{document}
\journalinfo{The Open Journal of Astrophysics}

\title{
A novel algorithm for GPU-accelerated particle-mesh interactions \\
implemented in the \textsc{Quokka} code
}

\author{
{Chong-Chong He$^{1}$\orcidlink{0000-0002-2332-8178}},
{Benjamin D.~Wibking$^{2}$},
{Aditi Vijayan$^{1}$},
{Mark R.~Krumholz$^{1}$\orcidlink{0000-0003-3893-854X}}, 
{Pak Shing~Li$^{3}$\orcidlink{0000-0001-8077-7095}}
}

\affiliation{
$^{1}$Research School of Astronomy and Astrophysics, Australian National University, 233 Mount Stromlo Road, Stromlo, ACT 2612, Australia\\
$^{2}$Department of Physics and Astronomy, Michigan State University, East Lansing, MI 48824, USA\\
$^{3}${Shanghai Astronomical Observatory,
Chinese Academy of Sciences, 80 Nandan Road, Shanghai 200030, PR China}
}

\email{chongchong.he@anu.edu.au}

\begin{abstract}
We present a novel, GPU-optimized algorithm for particle-mesh interactions in grid-based hydrodynamics simulations, designed for massively parallel architectures. This approach overcomes the inefficiency of particle neighbour searches or sorts across multiple GPU nodes by using a new ``particle-mesh-particle'' interaction scheme, which extends the particle-mesh method for self-gravity. The algorithm proceeds in two main stages: first, quantities exchanged between particles and the mesh -- such as mass, energy, and momentum added by stellar feedback or removed by accretion onto a sink -- are deposited into a buffer mesh equipped with ghost zones, where multiple contributions per cell are accumulated using atomic additions and then communicated across distributed memory ranks. In the second stage, the buffer states are applied to real mesh states, incorporating cell-wise limiters to enforce physical constraints such as positive density. We implement this scheme in the GPU-native radiation-magnetohydrodynamics code \textsc{Quokka} and validate it through a comprehensive suite of tests, including Bondi and Bondi-Hoyle accretion, and single and multiple supernova remnant evolution at varying spatial resolutions. We show that the algorithm achieves $\approx 50\%$ weak-scaling efficiency running on up to 8192 GPUs on the Frontier supercomputer. This scheme enables efficient, scalable particle-mesh coupling for GPU-optimized simulations of star formation and feedback in galaxies.
\end{abstract}


\maketitle

\section{Introduction}
\label{sec:intro}

Gravitational collapse is ubiquitous in gaseous astrophysical systems, but following collapsing regions in hydrodynamic simulations demands ever-increasing spatial and temporal resolution. It is therefore inevitable in such calculations that one must define a maximum resolution beyond which one does not follow further collapse. One standard approach to this problem is to replace unresolved high-density peaks with point masses that approximate the behaviour and evolution of unresolved small-scale structures, which may represent individual (proto-)stars, stellar populations, black holes, or similar compact objects. Subsequent to the creation of these particles, gas and particles are integrated simultaneously, including their mutual gravitational forces and possibly also interaction via other mechanisms such as radiation.

The integration of particles into hydrodynamic simulations has emerged as a vital approach in understanding complex astrophysical processes, from star formation and stellar feedback to black hole accretion.
\citet{Bate1995} first introduced the technique of sink particles -- particles that can accrete gas in addition to interacting with it gravitationally -- in smoothed particle hydrodynamics (SPH) codes, while \citet{Krumholz2004} pioneered their use in grid-based codes. Over the past two decades, similar implementations have appeared in various hydrodynamics codes, including \textsc{gadget} \citep{Jappsen2005}, \textsc{flash} \citep{Federrath2010}, \textsc{enzo} \citep{Wang2010}, \textsc{Ramses} \citep{Dubois2010,Bleuler2014}, and \textsc{Athena} \citep{Gong2013}. Most of these authors primarily focus on star formation simulations, but some also employ sink particles to model supermassive black holes in cosmological contexts.

In addition to accretion, particles are also frequently used as sources of feedback in hydrodynamic simulations. The oldest example of this is supernova feedback in simulations of galaxy formation, wherein gas above some density threshold is converted to collisionless particles that subsequently add energy back to the hydrodynamic particles or grid; the literature contains numerous recipes for treating this process \citep[e.g.,][]{Katz92a, Oppenheimer06a, Dalla-Vecchia12a, Kimm14a, Kim2017, Hopkins18c, Hirashima25a}. Other examples include particles representing stars that produce ionizing radiation \citep[e.g.,][]{Dale05a, Dale07c}, thermal radiation \citep[e.g.,][]{Krumholz07a, Krumholz09c, Offner09a}, protostellar jets \citep[e.g.,][]{Cunningham11a}, and massive stellar winds \citep[e.g.,][]{Dale14a, Gatto17a}, as well as particles representing black holes that similarly exert energetic and thermal feedback on the gas around them \citep[e.g.,][]{Springel05b, Di-Matteo05a}.

All of the algorithms above have been implemented in codes that run on CPU-based architectures, but these are now giving way to more powerful GPU-based systems that present particular programming challenges. Even on CPUs, neighbour searches often represent a major contributor to the computational cost in particle-based hydrodynamics methods, and this problem is worse on GPUs, where efficiency is limited by the data dependencies and irregular memory access patterns inherent to such methods. When running in parallel, the neighbour list generation step requires exchanging particles across all nodes, which becomes inefficient on multiple GPU nodes due to  complex synchronization and communication needs.
For instance, in GPU-accelerated SPH codes like that of \citet{Crespo2011plosone}, the calculation of interactions between a target particle and its neighbours represents the most computationally expensive step.

One might think that particle-mesh methods, in which one follows the hydrodynamics on a mesh while allowing particles to interact with that mesh, would avoid these challenges and allow efficient computation on GPUs. This is true to some extent. Meshes have the advantage of eliminating the neighbour search, since it is trivial to determine the computational cell in which a particular particle is located, and this in turn allows efficient deposition of particle information onto a grid, and conversely interpolation of mesh data back to particles; GPU-accelerated codes such as \textsc{Nyx} \citep{Almgren13a} and \textsc{WarpX} \citep{Fedeli22a} exploit this capability to achieve high performance for particle-mesh (PM) gravity and particle-in-cell (PIC) plasma simulations, respectively. However, in these applications particle-mesh interactions are far simpler than is the case for either sink accretion or stellar feedback; particle deposition to the mesh in both PM and PIC simulations is simply additive, but this is \textit{not} the case for either accretion or feedback, which usually involve complex particle-mesh interaction rules that are problematic to implement on GPU.

As one example of this difficulty, most recent implementations of supernova feedback \citep{Kim2017, Hopkins18c, Hirashima25a} involve modifying the properties of cells or particles within some distance of the explosion site with recipes such that the results when multiple supernovae occur are strongly dependent on the order in which the particles contributing supernovae are processed; that is, the results of adding supernova feedback from particle A and then B are not the same as the result of doing so from B and then A. On a CPU one can resolve this issue simply by enforcing an order in which the particles are processed, but ensuring serially-ordered processing of particles that are potentially distributed across multiple nodes in memory is either impossible or prohibitively expensive on GPUs. To make matters worse, if multiple GPUs need to process the supernova feedback recipe (for example because the supernovae are occurring at the edge of a domain decomposition), there is no easy way to guarantee that they will do so in the same order, and thus a naive implementation of CPU-based supernova feedback recipes might yield inconsistent results from one GPU to another. This is just one example of the numerous rules for particle merging, sorting, or prioritization invoked by the particle recipes listed above that do not translate trivially to GPU architectures.

To overcome these problems and allow efficient calculation of more complex particle-mesh interactions on modern GPU hardware, in this paper we introduce a novel ``particle-mesh-particle'' algorithm. This method specifically tackles the challenges of particle-particle and particle-mesh interactions in a GPU-friendly framework. The major advantages of this algorithm include avoiding the need for hard-to-accelerate particle neighbour searches across distributed GPU memory, producing communication and computation in a predictable pattern with no indirect memory referencing to enable efficient GPU acceleration, and requiring only a single communication step per update cycle. We implement this method in the \quokka~GPU-accelerated radiation-hydrodynamics code \citep{Wibking2022, He2024a, He2024b}.

The remainder of this paper is organized as follows. In \autoref{sec:alg}, we describe the overall algorithm and communication protocols for particle-mesh and particle-particle interactions. In \autoref{sec:impl}, we detail our implementations of sink particle formation and accretion (appropriate for simulations of individual star formation) and stellar population formation and supernova feedback (appropriate for galaxy-scale simulations) within this framework, and in \autoref{sec:test} we present a series of tests that demonstrate the accuracy and efficiency of this approach. Finally, \autoref{sec:summary} summarizes our findings.

\section{The ``particle-mesh-particle'' algorithm}
\label{sec:alg}

\subsection{Formulation of the problem}

In this section, we outline the fundamental algorithm that we use for particle-mesh interactions. We consider a mesh consisting of quadrilateral cells with indices $ijk$, each of which stores a vector of conserved quantities $\vecU_{ijk}$. In a minimal hydrodynamics application $\vecU_{ijk}$ would contain the mass density, momentum density, and total energy in each cell, but this can be straightforwardly generalized to more complex methods where $\vecU_{ijk}$ might include, for example, mass densities of particular chemical species or radiation energy densities and fluxes; we require only that the quantities stored in $\vecU_{ijk}$ be conserved volumetric quantities. Similarly, the mesh might be uniformly-spaced, or it could be an adaptive mesh in which the cell volume $V_{ijk}$ varies from cell to cell; this too will not matter for our purposes. Finally, the mesh may be domain-decomposed in memory, so that any given GPU only has access to some fraction of the cells making up the computational domain. 

Within this mesh we also have a series of particles, indexed by $s$; each particle is characterised by a position $\vecx_s$, which lies within some host cell $\{ijk\}_s$. Particles follow the same domain decomposition as the mesh, i.e., if cell $\{ijk\}_s$ lies within the domain of a given GPU, then particle $s$ will also be stored on that GPU. Particles may also carry conserved quantities $\vecu_s$ that correspond to the quantities stored in the state variable and obey conservation laws; for example, sink particles typically carry a total mass and momentum, and the accretion interaction between particles and gas must conserve total mass and momentum, so
\begin{equation}
    \sum_{ijk} \vecU_{ijk} V_{ijk} + \sum_s \vecu_s = \mathrm{constant},
    \label{eq:conservation}
\end{equation}
where $\vecu = (m, m\mathbf{v})$ are the mass and momentum of a particle with velocity $\mathbf{v}$, and $\vecU = (\rho, \vecp)$ are the cell mass and momentum density.

Now consider an interaction whereby particle $s$ should alter the conserved quantities in cell $ijk$ by an amount $\Delta \vecU^s_{ijk}$; we refer to  $\Delta \vecU^s_{ijk}$ as the deposition function, which specifies the amount of each conserved quantity that a particle deposits in each cell. Deposition can take the form of a direct and permanent alteration to the conserved quantities themselves (for example accretion of mass by a sink particle), or to a temporary deposition for the purposes of a subsequent calculation (for example depositing particle mass into the gas density field for the purposes of computing the gravitational potential as in a PM gravity method), and can be both positive (for example addition of energy by SN feedback) or negative (for example reduction in density due to sink accretion). Particles can affect cells other than their host cell, but we require that $\Delta \vecU^s_{ijk}$ be non-zero only over a kernel of radius $r_K$ centred on the particle position, so that $\Delta\vecU^s_{ijk}$ is zero for any cell whose central position $\vecx_{ijk}$ satisfies $|\vecx_{ijk} - \vecx_s| > r_K$. 

Crucially, we allow multiple particles to impose alterations on a single cell, i.e., we can have $\Delta \vecU^s_{ijk} \neq 0$ and $\Delta \vecU^{s'}_{ijk} \neq 0$ for two distinct particles $s$ and $s'$, and we allow limiting, so that if $\sum_s \vecU^s_{ijk}$ is too large -- for example if the collective effect of multiple sink particles accreting from a given cell would be to make its density negative, or the collective effect of momentum addition from multiple supernovae would lead to a violation of the total supernova energy budget -- we can modify the sum before altering the conserved quantities. These two features are important for GPU-based implementations, because without them we would be forced to implement an algorithm to find neighbouring particles whose deposition zones overlap and implement logic for handling that case (e.g., by merging particles), and it is precisely such an operation that we wish to avoid because it is inefficient on GPU.

\subsection{Algorithm steps}
\label{ssec:alg_steps}

\begin{figure*}
    \includegraphics[width=\textwidth]{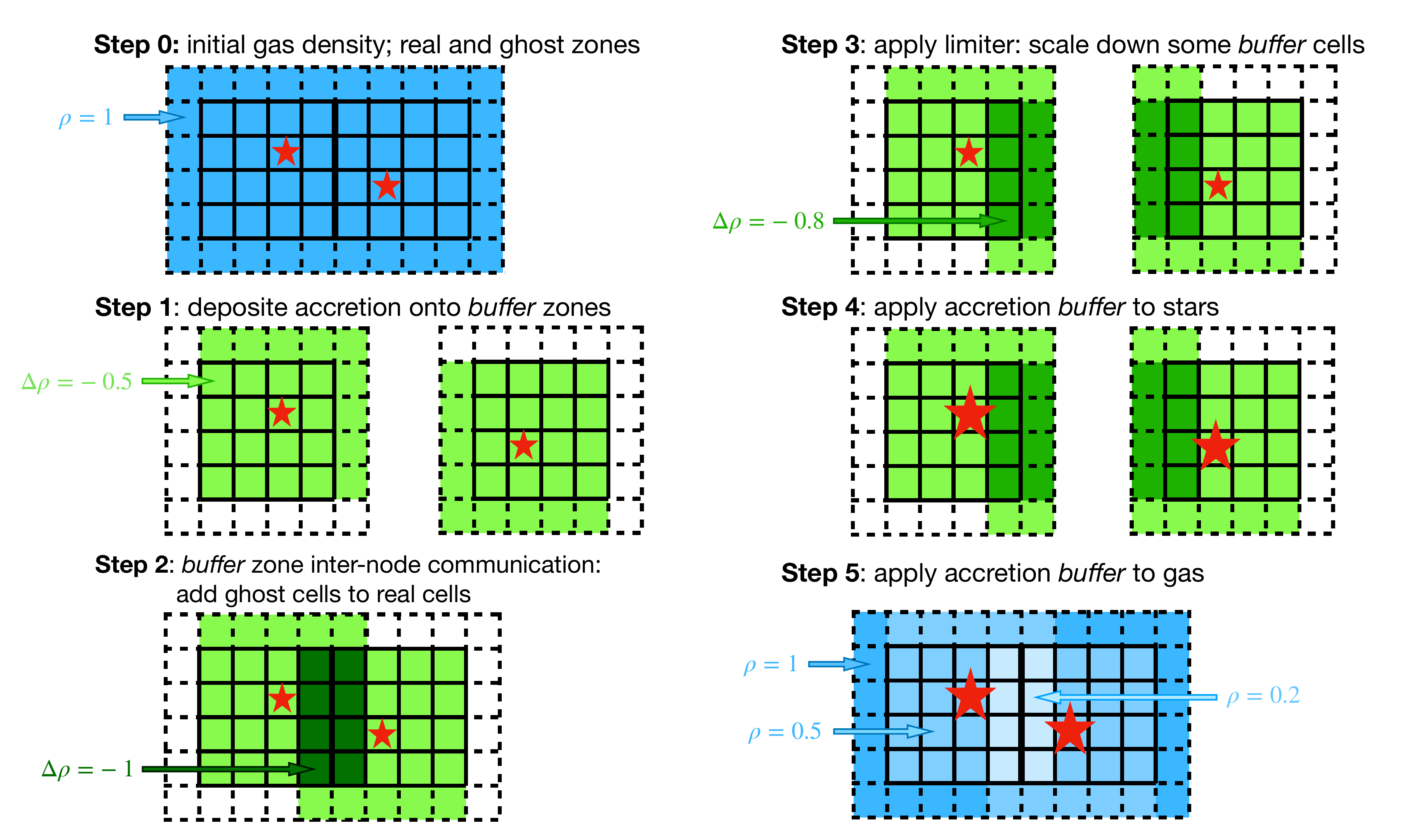}
    \caption{\edit{Summary of the particle-mesh-particle algorithm for particle-gas interactions on GPUs, illustrated for the case of sink accretion. In the example, two sink particles (red stars) lie near the boundaries between two grid blocks that reside on different GPUs; the thick black line indicates the block boundary. Cells with dashed edges are ghost zones used for communication. For clarity we show only one layer of ghost zones and two-cell deposition kernel in this schematic, but our production runs employ four or more ghost zones and three-cell deposition kernel. Blue shading indicates gas density, while green shows the buffer quantity (here, the accretion rate). The annotated numbers show example values at various steps: the initial density $\rho = 1$ at Step 0, the change in density $\Delta\rho = -0.5$ computed independently over the accretion kernel for the two particle on each rank at Step 1, the inter-rank buffer sum that increases the mass change to $\Delta\rho = -1$ in the overlapping cells at Step 2, the limiter that scales down this change to $\Delta\rho = -0.8$ at Step 3, the increase in sink mass at Step 4, and the decrease in gas density to new, smaller values in Step 5.}
    \label{fig:pmp}
    }
\end{figure*}

With this general framework in place, we now describe our algorithm, emphasizing the synchronization and communication protocols critical for GPU optimization. We defer specific details of how we implement different particle types to \autoref{sec:impl}, and we provide additional technical details of our method to achieve bitwise-reproducible results in \aref{app:reproducibility}. The algorithm comprises five steps, which we summarize in \autoref{fig:pmp}:\\

\textbf{Step 1: Particle to buffer mesh.} First, we allocate a buffer of conserved quantities $\Delta\vecU_{ijk}^\mathrm{buf}$ with the same dimensions as the hydrodynamic mesh, with all cells initialized to zero. In a distributed-memory parallel calculation this buffer is domain-decomposed exactly as is the hydrodynamic mesh, but it includes enough ghost zones (also known as guard cells) to fully contain the interaction kernel radius $r_K$, so if the host cell $\{ijk\}_s$ of particle $s$ is owned by a given GPU, then all the cells $ijk$ for which $\Delta \vecU_{ijk}^s$ is non-zero are included in the buffer and its ghost zones. We then iterate over all particles and evaluate the deposition function $\Delta\vecU_{ijk}^s$ for all cells and all particles in local memory, storing the result to the buffer, i.e., we carry out the operation
\begin{equation}
    \Delta\vecU_{ijk}^\mathrm{buf} = \sum_s \Delta \vecU_{ijk}^s,
    \label{eq:delta_u_buf}
\end{equation}
where the sum runs over all particles $s$ stored on a given GPU. This deposition must be carried out with some care to avoid race conditions on GPU, as we discuss further in \aref{app:reproducibility}.\\

\textbf{Step 2: Inter-GPU summation of the buffer mesh.} We next sum the buffer mesh across distributed memory ranks, i.e., we add the contents of the ghost zones on each rank to the corresponding real zones on other ranks, so that at the end of this procedure $\Delta\vecU_{ijk}^\mathrm{buf}$ contains the contributions from all particles $s$ that deposit to a given cell, even if those particles are hosted on a different rank. We denote these parallel-summed quantities
\begin{equation}
    \Delta\vecU_{ijk}^\mathrm{sum} = \sum_s \Delta \vecU_{ijk}^s,
\end{equation}
where this expression differs from \autoref{eq:delta_u_buf} in that this summation is over all particles $s$ anywhere in the computational domain, not just those that reside on the same GPU as the cell. Again, some care is required in carrying out this step, as described in \aref{app:reproducibility}.\\

\textbf{Step 3: Apply limiter.} As noted above, to avoid unphysical states when multiple particles deposit to the same hydrodynamic cell, it may be necessary to apply limiters -- for example limiting the amount of mass removed from a given cell by sink accretion to avoid making the density negative. We defer discussion of the specific forms of limiting we implement for different particle-mesh interaction types to \autoref{sec:impl}, and here simply abstract the limiting process as a cell-wise operation of the form
\begin{equation}
    \Delta\vecU^\mathrm{lim} = f_\mathrm{lim}(\vecU, \Delta\vecU^\mathrm{sum}),
\end{equation}
where $f_\mathrm{lim}$ is some limiter. We store the ratio of the limited and unlimited quantities $\boldsymbol{\eta}_{ijk} = \Delta\vecU_{ijk}^\mathrm{lim}/\Delta\vecU_{ijk}^\mathrm{sum}$ for use in the next steps. Note that $\boldsymbol{\eta}_{ijk}$ is a vector that can be different for each conserved quantity, though in practice it is often the same for all quantities.\\

\textbf{Step 4: Apply limited changes to particles.} The fourth step is to update particle properties to account for particle-mesh interaction; note that for particles we carry out this update here, rather than as part of step 1, because limiting may affect the update to particle properties -- again returning to the example of sink particles, if we have limited the change in density $\Delta\rho_{ijk}$ in some cell $ijk$, then the change in mass of the sink particles that accrete from this cell must be similarly affected so as to ensure conservation of total mass. Thus in this step we repeat the calculation of $\Delta \vecU_{ijk}^s$ for each particle from step 1, summing over all the cells $ijk$, and we then update the particle properties based on the limited changes in each cell:
\begin{equation}
    \vecu_s := \vecu_s - \sum_{ijk} \boldsymbol{\eta}_{ijk} \Delta\vecU_{ijk}^s V_{ijk}.
\end{equation}

\textbf{Step 5: Apply limited changes to the mesh.} The final step is to deposit the limited quantities back onto the mesh. Mathematically, this step amounts to carrying out
\begin{equation}
    \vecU_{ijk} := \vecU_{ijk} + \Delta\vecU_{ijk}^\mathrm{lim} =  \vecU_{ijk} + \boldsymbol{\eta}_{ijk} \sum_s \Delta\vecU_{ijk}^s
\end{equation}
over all cells. It is manifestly clear that this process obeys the conservation constraint \autoref{eq:conservation}. \\

Before moving on to the details of how we implement this approach for both sink particle accretion and supernova feedback in \autoref{sec:impl}, we pause to make some observations. First, not all steps of this algorithm are required for all types of particles. Simple particle-mesh interactions of the type used by PM or PIC simulations only require steps 1, 2, and 5, because there is no limiting and because interactions do not modify particle properties, only mesh properties. Similarly, some of the particle algorithms we describe in \autoref{sec:impl} will use only some of the steps.

Second, the only step in the entire algorithm that is not local to a single GPU is step 2, and the communication pattern associated with this step is \textit{exactly} the same as that used for ghost zone filling during the hydrodynamic update. Thus as long as the interaction kernel radius $r_K$ is small enough that interaction buffers do not require more ghost zones than hydrodynamics, the computational cost of this step is guaranteed to be no larger than the computational cost of hydrodynamic ghost cell filling. This cost is minimized by the usual approach of using a space-filling curve to carry out domain decomposition such that physically-adjacent parts of the domain are whenever possible hosted on the same computational node, and this strategy is sufficiently efficient that GPU-optimized hydrodynamics codes such as \quokka~and \textsc{AthenaK} \citep{Stone24a} show near-linear weak scaling even when running on tens of thousands of GPUs.

\section{Numerical implementations: sink and supernova feedback particles}
\label{sec:impl}

Having outlined our general particle-mesh-particle interaction framework, we now provide detailed example implementations for two important use-cases: sink particles for star formation simulations (\autoref{ssec:sink}) and stellar population particles with supernova feedback for galaxy formation simulations (\autoref{ssec:sn}).

\subsection{Sink particles}
\label{ssec:sink}

As described in \autoref{sec:intro}, sink particles are commonly used to represent accreting protostars in simulations of star formation, and thus the key interaction between such particles and the hydrodynamic mesh that we seek to capture is accretion. Our implementation of sink particles in \quokka{} carries out the following steps during every time step on the finest adaptive mesh refinement level:
\begin{enumerate}
  \item Integrate the motion of existing sink particles. We carry out this step using the leapfrog method to integrate the particle position and velocities in time, including the effects of the gravitational interactions between sink particles and gas; we compute the gravitational accelerations using a standard PM gravity method, which we implement using steps 1, 2, and 5 of the algorithm described in \autoref{sec:alg}. We follow a ``kick-drift-kick'' strategy, which is shown to work better than ``drift-kick-drift'' approach for variable time step \citep{Springel2005a}. 
  \item Compute accretion onto existing sink particles using all steps of the algorithm in \autoref{sec:alg}; we discuss the implementation details in  \autoref{sec:acc}. We retain the buffer mesh containing the cell-wise accretion rates at the end of this step, for use in the subsequent step.
  \item Create sink particles following the algorithm outlined in \autoref{sec:sinkc}; this takes the buffer mesh constructed during the accretion stage as an auxiliary input. 
\end{enumerate}
The details of the accretion and particle creation steps are as follows.

\subsubsection{Sink particle accretion}
\label{sec:acc}

Accretion from sink particles uses all steps of the algorithm outlined in \autoref{sec:alg}; to fully specify our implementation, we must therefore define our kernel radius $r_K$ and provide the functions $\Delta \vecU_{ijk}^s$, which describes the change in conserved quantities in each cell due to accretion onto each sink, and $f_\mathrm{lim}(\vecU,\Delta\vecU)$, the limiter function.

Our default choice is $r_{K} = 3 \Delta x$, where $\Delta x$ is the cell spacing, but this is adjustable at compile time, and our tests below show similar results for $r_K = 2 \Delta x$ or $4 \Delta x$ (see \autoref{ssec:bondi}). We take our model for the accretion function $\Delta\vecU_{ijk}^s$ from \citet{Krumholz2004}, who use the Bondi-Hoyle accretion formula to approximate the accretion rate. In this approach, we first compute the Bondi-Hoyle radius for each particle
\begin{equation}\label{eq:rbh}
    r_{\mathrm{BH}}=\frac{G m_s}{v_{\infty}^2+c_{\infty}^2},
\end{equation}
where $m_s$ is the sink mass, and $v_{\infty}$ and $c_{\infty}$ are the mass-weighted mean relative velocity and mass-weighted mean sound speed over cells within the accretion kernel, respectively. Next, we compute the Bondi-Hoyle-Lyttleton accretion rate \citep{Hoyle1939}, using the approximation provided by \citet{Ruffert1994} to interpolate between the limits of $v_{\infty} \ll c_{\infty}$ and $v_{\infty} \gg c_{\infty}$:
\begin{equation}\label{eq:mdot}
\begin{aligned}
    \dot{m}_\mathrm{BH}&=4 \pi \rho_{\infty} G^2 m_s^2\left[\frac{\lambda^2 c_{\infty}^2+v_{\infty}^2}{\left(c_{\infty}^2+v_{\infty}^2\right)^4}\right]^{1 / 2} \\
    &=4 \pi \rho_{\infty} r_{\mathrm{BH}}^2\left(\lambda^2 c_{\infty}^2+v_{\infty}^2\right)^{1 / 2}
\end{aligned}
\end{equation}
where $\lambda$ is a constant of order unity that depends on the gas equation of state; we adopt the value for isothermal gas, $\lambda = e^{3/2}/4 \approx 1.120$, throughout this work. Third, we apply an accretion kernel to distribute the accretion rate across the accretion zone, weighting each cell by
\begin{equation}\label{eq:kernel}
    w = \exp \left(-r^2 / r_\mathrm{acc}^2\right),
\end{equation}
where $r = |\mathbf{x}_s - \mathbf{x}_{ijk}|$ is the distance from the particle to the cell centre, and
\begin{equation}
    r_\mathrm{acc} = \begin{cases}\Delta x / 4, & r_{\mathrm{BH}}<\Delta x / 4 \\ r_{\mathrm{BH}}, & \Delta x / 4 \leq r_{\mathrm{BH}} \leq r_K / 2 \\ r_K / 2, & r_{\mathrm{BH}}>r_K / 2\end{cases}.
\end{equation}
Thus the normalised weight for each cell inside the accretion kernel is
\begin{equation}
    \edit{\phi_{ijk} = \frac{w_{ijk}}{\sum w_{ijk}},}
    \label{eq:kernel_wgt}
\end{equation}
where the sum runs over all cells inside the accretion kernel. Fourth, we increase the accretion rate in any cell where the density is high enough to violate the \citet{Truelove1997} condition for stability against artificial fragmentation, which requires that the density not exceed
\begin{equation}
    \rho_\mathrm{Tr} = J^2 \frac{\pi c_s^2}{G\Delta x^2}.
    \label{eq:truelove}
\end{equation}
Here $J=0.25$ is the Jeans number required for stability, $c_s$ is the sound speed in the cell, and $\Delta x$ is the cell size. In cells for which $\rho_{ijk} > \rho_\mathrm{Tr}$, we increase the accretion rate enough to ensure that the density falls to $\rho_\mathrm{Tr}$, thereby ensuring that the calculation does not undergo artificial fragmentation. Putting these conditions together, our final expression for the accretion function for minimal hydrodynamics (for which the conserved quantities are density, momentum, and total energy) is
\begin{equation}
    \Delta\vecU_{ijk}^s = \max\left(\dot{m}_\mathrm{BH},\dot{m}_\mathrm{Tr}\right)\frac{\phi_{ijk}\Delta t}{V_{ijk}} \left(
    \begin{array}{c}
        1\\
        \mathbf{v} \\
        e_\mathrm{sp} + v^2/2
    \end{array}
    \right)_{ijk},
    \label{eq:accretion_function}
\end{equation}
\edit{where $\phi_{ijk}$ is the kernel weight (\autoref{eq:kernel_wgt})} and
\begin{equation}
    \dot{m}_\mathrm{Tr} = \frac{(\rho_{ijk}-\rho_\mathrm{Tr}) V_{ijk}}{\phi_{ijk}\Delta t},
\end{equation}
is the accretion rate required to enforce the \citeauthor{Truelove1997} condition, $\Delta t$ is the time step, $V_{ijk}$ is the cell volume, and $\mathbf{v}$ and $e_\mathrm{sp}$ are the cell velocity and specific internal energy. Extensions beyond minimal hydrodynamics are straightforward. \quokka~includes a dual-energy formalism to follow high-Mach number flows, and so we also follow the internal energy as an additional, auxiliary conserved quantity; the accretion formula for this quantity is the same as for total energy, but with $e_\mathrm{sp} + v^2/2$ replaced by $e_\mathrm{int,sp}$, where $e_\mathrm{int,sp}$ is the internal energy per unit mass. In calculations including chemistry or passive scalars, we set the changes in their densities to keep their concentrations invariant, i.e., we choose $\Delta\rho_\mathrm{sp} = (\rho_\mathrm{sp}/\rho)\Delta \rho$, where $\rho_\mathrm{sp}$ is the density of a given species. Finally, in magnetohydrodynamic calculations, we follow \citet{Myers13a} by not altering magnetic fields or magnetic energy densities to avoid artificially creating magnetic divergence.

Our limiter function is chosen to ensure stability by ensuring that the mass in a given cell is reduced by no more than 25\% per time step \citep{Krumholz2004} unless doing so is necessary to enforce the \citeauthor{Truelove1997} condition. If the mass accreted is limited, then the momentum and energy changes are limited by the same factor. Thus our limiting function is
\begin{equation}
    f_\mathrm{lim}(\vecU, \Delta\vecU^\mathrm{sum}) = \left(\frac{\Delta\rho^\mathrm{lim}}{\Delta\rho}\right) \Delta\vecU,
\end{equation}
where
\begin{equation}
    \Delta\rho^\mathrm{lim} = \max\left[\min\left(\Delta\rho, \frac{\rho}{4}\right), \rho - \rho_\mathrm{Tr}\right].
\end{equation}
Equivalently, the ratio of the limited to unlimited changes is $\boldsymbol{\eta}_{ijk} = \Delta\rho^\mathrm{lim}/\Delta\rho$ for all quantities.

\subsubsection{Sink particle creation}
\label{sec:sinkc}

We also follow a slightly-modified version of the \citet{Krumholz2004} criterion for sink particle creation, which is to insert a sink particle in cells that violate the \citet{Truelove1997} condition for artificial self-gravitating fragmentation. This condition amounts to requiring that the ratio of the cell size to the local Jeans length remain below a threshold $J\approx 0.25$, which in turn yields the density threshold given by \autoref{eq:truelove}. We therefore mark cells within which $\rho_{ijk} > \rho_\mathrm{Tr}$ as candidates for sink particle creation. However, we only create sink particles in cells that satisfy two additional criteria that were not present in the original \citeauthor{Krumholz2004} formulation: we require that a sink particle is only created if: (1) the cell is a local density maximum within a sphere of radius $r_K$; (2) no other sink particle exists within $r_K$.\footnote{We do not add the further restrictions on sink particle formation (e.g., requiring $\nabla\cdot\mathbf{v} < 0$) advocated by some authors \citep[e.g.,][]{Federrath2010, Haugbolle18a}. We do not do so because these criteria substantially affect the outcome of fragmentation only when the gas remains isothermal or close to it up to the sink particle creation density; if the effective equation of state stiffens at lower densities, either due to artificial stiffening \citep[e.g.,][]{Jappsen2005, Kratter10a} or inclusion of physical processes such as radiative transfer \citep[e.g.,][]{Krumholz07a, Krumholz09c}, fragmentation occurs on resolved scales and the results become relatively insensitive to the details of the sink particle prescription. In the isothermal limit where details of the sink particle creation criteria do matter, it is far from clear that it is possible or meaningful to obtain a converged result for fragmentation \citep[e.g.,][]{Guszejnov18a}. For this reason, we do not attempt to implement more involved sink particle creation recipes.}

We enforce the latter condition by forbidding sink particle creation in any cell within which $\Delta\vecU_{ijk}$ is non-zero. We add these two additional criteria due to a subtle change required in the \citeauthor{Krumholz2004} approach to make it GPU-friendly. In the original approach, sink particles could be created in multiple adjacent cells, but were then immediately merged, which prevented the creation of large numbers of sink particles in close proximity. We do not use this approach here because selecting particles to merge requires constructing particle trees across distributed GPU memory, an inefficient operation that we wish to avoid. Our approach of only allowing sink creation at local maxima and not within the accretion kernel of another sink particle, but then increasing the accretion rate itself enough to enforce the \citeauthor{Truelove1997} condition, achieves the same effect as the original \citeauthor{Krumholz2004} prescription without the additional computational overhead of constructing parallel particle trees on GPU.

If a cell does satisfy the sink creation conditions, we set the initial sink mass to $m_s = (\rho - \rho_{\rm Tr}) V_{ijk}$, and reduce the cell density to $\rho_{\rm Tr}$. We leave the cell velocity and specific internal and kinetic energies unchanged, and set the initial momentum of the sink particle to $m_s \mathbf{v}_{ijk}$, where $\mathbf{v}_{ijk}$ is the velocity of the cell in which the sink particle is created.

\subsection{Massive stellar particles and supernova feedback}
\label{ssec:sn}

In simulations of entire galaxies, the resolution is generally too low to capture individual accreting protostars, which begin their accretion upon second collapse after reaching masses of only $\sim 10^{-3}$ M$_\odot$. Instead, the particles used in galaxy-scale simulations are intended to represent stellar populations. Depending on the resolution, such populations can be either integrated, representing a collection of stars large enough to treat supernovae as a continuous wind, or stochastic, representing stellar populations small enough to treat each supernova individually. The implementation for GPU we present here is an example of the latter, and thus is similar in spirit to the stochastic stellar population models implemented by, for example, \citet{Gatto17a}, \citet{Fujimoto18a}, \citet{Armillotta19a}, \citet{Applebaum20a}, and \citet{Jeffreson21b} for the \textsc{flash}, \textsc{enzo}, \textsc{gizmo}, \textsc{changa}, and \textsc{arepo} codes, respectively.

Our basic steps for supernova feedback particles are much the same as for sink particles. During each time step, we
\begin{enumerate}
    \item Integrate the motion of existing particles under the influence of gravity using a ``kick-drift-kick'' integration, with the potential computed using a PM method.
    \item Inject supernova feedback from stars at the ends of their lives (\autoref{sssec:sne}).
    \item Create new particles in cells where the density has risen past our ability to resolve (\autoref{sec:sto}).
\end{enumerate}
We describe the second and third steps in this algorithm in detail in the next two sections.

\subsubsection{Supernova feedback}
\label{sssec:sne}

As we discuss further in \autoref{sec:sto}, each of our particles represents a single massive star, which has a known formation time and an initial mass assigned at birth. We determine the lifetime of each such star by interpolating on the MIST tracks for rotating, Solar-metallicity stars \citep{Choi16a}, and we determine which stars end their lives in core collapse supernovae from the tabulation of \citet{Sukhbold16a}. For each time step we first find all the stars whose lives end during that time step, and then inject supernova back from those stars. Feedback injection follows the general scheme laid out in \autoref{sec:alg}, with the exception that we skip step 4 because there is no feedback from the computational grid to the properties of the stellar particles. The remaining steps can be fully specified by the choice of accretion kernel radius $r_K$ (as with sink particles, we use $r_K = 3\Delta x$ by default), the deposition function $\Delta\vecU_{ijk}^s$, and the limiter $f_\mathrm{lim}(\vecU, \Delta\vecU)$.

\paragraph{The deposition function.}
We calculate the deposition function using a variant of the TIGRESS prescription \citep{Kim2017} adapted to be GPU-friendly.
Following the TIGRESS approach, we first calculate the total gas mass within the kernel radius $r_K$, which we take to be $M_\mathrm{snr} = \sum \omega_{ijk} \rho_{ijk} V_{ijk} + M_\mathrm{ej}$, where the sum runs over all cells within the stencil and $\omega_{ijk}$ is the fraction of the grid cell that is covered by the stencil, and $M_\mathrm{ej}$ is the mass ejected by the supernova itself. In a production simulation we obtain $M_\mathrm{ej}$ by interpolating on the tabulated values provided by \citet{Sukhbold16a} in the same way we determine which stars end their lives as supernovae, but for the purposes of comparing with the results of \citeauthor{Kim2017} we follow them in setting $M_\mathrm{ej} = 10$ M$_\odot$ for all stars in the tests presented below. We then compute the ratio of this mass to the shell-formation mass, defined as the approximate swept-up mass at the point when the interior of a supernova remnant transitions from adiabatic to radiative and thus forms a dense shell. This ratio is \(\mathcal{R}_M \equiv M_{\rm snr} / M_{\rm sf}\), where \(M_{\rm sf} = 1679 (n_{\rm H} / \mathrm{cm}^{-3})^{-0.26}\) M$_\odot$ and $n_\mathrm{H}$ is the ambient number density of H nuclei \citep{Kim2015}. We estimate the latter quantity as $n_\mathrm{H} = M_\mathrm{snr} / \mu_\mathrm{H} V_\mathrm{snr}$, where $\mu_\mathrm{H}$ is the mass per H nucleus ($\approx 1.4 m_\mathrm{H}$ for standard He abundances) and $V_\mathrm{snr} = \sum \Delta V_{ijk}$ is the total volume of the accretion kernel; here the sum again goes over cells satisfying $|\mathbf{x}_{ijk} - \mathbf{x}_s| < r_K$.

We then calculate the deposition function using three different prescriptions depending on the value of \(\mathcal{R}_M\); following \citeauthor{Kim2017}, we refer to these as the \texttt{EJ} (Ejecta), \texttt{ST} (Sedov-Taylor), and \texttt{MC} (Momentum Conserving) regimes, depending on whether the resolution is sufficient to capture the transition between free expansion of the ejecta and the adiabatic Sedov-Taylor phase (the \texttt{EJ} case), the transition between the Sedov-Taylor phase and the momentum-conserving snowplow phase (the \texttt{ST} case), or neither of these transitions (the \texttt{MC} case). 
We describe our deposition function with the state array (with mass, momentum, and total energy as conserved quantities)
\begin{equation}
\label{eq:du}
    \Delta\vecU_{ijk}^s =
    \left(
    \begin{array}{c}
    \omega_{ijk} M_\mathrm{ej} / V_\mathrm{snr} \\
    \rho_{\rm new} \vec{v}_{\mathrm{COM}} - \rho_{ijk} \vec{v}_{ijk} + \vec{p}_{\mathrm{radial}} \\
    \omega_{ijk} \left(E_{\mathrm{sn}} + E_{\mathrm{kin}}\right) / V_{\mathrm{snr}} + \vec{v}_{\text{COM}} \cdot \vec{p}_{\mathrm{radial}}
    \end{array}
    \right),
\end{equation}
where $\rho_{\rm new} = \rho_{ijk} + \omega_{ijk} M_\mathrm{ej} / V_\mathrm{snr}$ is the updated cell density, $E_\mathrm{sn} = 10^{51}$ erg is the supernova energy, $E_{\text{kin}} = \frac{1}{2} m_{\text{ej}} v_s^2$ is the kinetic energy of the supernova ejecta, $\vec{p}_{\mathrm{radial}} = |\vec{p}_{\mathrm{radial}}| \hat{r}$ is the supernova momentum injection in the radial direction in a coordinate system centred on the exploding particle, and $\vec{v}_{\text{COM}}$ is the centre-of-mass velocity of the supernova remnant,
\begin{equation}
    \vec{v}_{\text{COM}} = \frac{\sum \omega_{ijk} \rho_{ijk} V_{ijk} \vec{v}_{ijk} + m_{\text{ej}} \vec{v}_{\text{ej}}}{M_{\text{snr}}}.
\end{equation}

The cross term $\vec{v}_{\text{COM}} \cdot \vec{p}_{\text{radial}}$ accounts for the work done by the radial expansion against the bulk motion of the remnant. This formulation effectively smooths the velocity field (but not the density) before deposition. Because $\vec{p}_{\text{radial}}$ is isotropic, the cross term cancels when summing over all cells in the stencil, and thus the work done on all cells by the radial expansion sums to zero, ensuring global energy conservation. This also guarantees the supernova deposition is Galilean invariant, i.e., the outcome of a SN event is independent of the velocity of the frame in which it is computed.

To motivate this procedure, it is helpful to evaluate what occurs if we do not smooth the background velocity and only add the radial component, i.e.\ $\Delta p = \vec{p}_{\text{radial}}$. In this case, the cross term in the energy update becomes $\vec{v}_{ijk} \cdot \vec{p}_{\text{radial}}$, which, when summed over the stencil, becomes $\sum \vec{v}_{ijk} \cdot \vec{p}_{\text{radial}} = \sum (\vec{v}_{\text{COM}} + \delta \vec{v}_{ijk}) \cdot \vec{p}_{\text{radial}} = \sum \delta \vec{v}_{ijk} \cdot \vec{p}_{\text{radial}}$, where $\delta \vec{v}_{ijk}$ is the cell velocity relative to the centre of mass. This introduces spurious energy into the system, which can be significant when a supernova event occurs shortly after another at the same location, such that $\delta \vec{v}_{ijk}$ is parallel to $\hat{r}$ and thus $\vec{p}_{\text{radial}}$ in all cells. We note that, by smoothing the background velocity, the kinetic energy in the centre-of-mass frame before deposition is lost. However, because we update the total energy and the internal energy is derived from the other conserved quantities, this lost energy effectively becomes internal energy, which is subsequently handled by radiative cooling.

The magnitude of the momentum deposition $|\vec{p}_{\text{radial}}|$ changes between regimes as
\begin{eqnarray}
\label{eq:dp}
    |\vec{p}_{\text{radial}}| & = & \begin{cases}
        P_\mathrm{snr}, & \mathcal{R}_M > 1 \,(\texttt{MC}) \\
        \sqrt{2 M_\mathrm{ej} \epsilon_K E_{\rm sn}}, & 0.027 < \mathcal{R}_M \leq 1 \,(\texttt{ST}) \\
        0, & \mathcal{R}_M \leq 0.027 \,(\texttt{EJ})
    \end{cases}
\end{eqnarray}
Here $\epsilon_K = 0.28$ is the fraction of energy in kinetic form during the Sedov-Taylor phase and $P_\mathrm{snr}$
is the terminal momentum we expect at the end of the Sedov-Taylor phase. These expressions amount to assuming that in the \texttt{MC} case the supernova adds only kinetic energy and momentum and so the internal energy per unit mass stays constant, while in the \texttt{ST} regime the changes in kinetic and internal energy match those expected for the Sedov-Taylor solution, and in the \texttt{EJ} case the supernova adds only thermal energy. Note that, because we are always depositing an energy of $E_{\text{sn}}$ into the gas total energy, in the \texttt{MC} case the supernova adds a small amount of thermal energy to the gas because the total kinetic energy deposited is $P_{\text{snr}}^2 / 2 M_{\text{ej}} = 4.6 \times 10^{50}$ erg $= 0.46 E_{\text{sn}}$. The remaining 54 per cent of the energy is deposited as thermal energy. This assignment of thermal and kinetic energy is handled with the limiter step discussed below.

We adopt a numerical value
\begin{equation}
    \label{eq:psnr}
    P_\mathrm{snr} = 2.8 \times 10^5 \, \mathrm{M}_{\odot} \, \mathrm{km} \, \mathrm{s}^{-1} \left(\frac{n_{\mathrm{H}}}{\mathrm{cm}^{-3}}\right)^{-0.17}
\end{equation}
for the terminal momentum directly from the \citeauthor{Kim2017} prescription. This is a fit to results from the simulations of \citet{Kim2015}; however, this numerical value is roughly consistent with those produced by other studies of the remnants of single, isolated supernovae in both one dimension \citep{Blondin98a, Thornton98a} and three dimensions \citep{Iffrig15a, Martizzi15a}.\footnote{The terminal momentum of the remnants produced by multiple supernova clustered closely enough in time that several explosions contribute to the expansion remains significantly uncertain, and may well depend on the degree of inhomogeneity or other properties of the background into which those remnants expand; see \citet{Gentry17a, Gentry19a, Gentry20a}, \citet{Kim17a}, and \citet{El-Badry19a} for further discussion.}

\paragraph{The limiter.}
Our remaining task is to specify the limiter $f_\mathrm{lim}(\vecU, \Delta\vecU)$. Limiting is required because \autoref{eq:du} does not guarantee that the state produced after deposition obeys the constraint that $e - |\mathbf{p}|^2/2\rho \ge e_{\rm int, 0}$, where $e$ and $\bp$ are the total energy and momentum per unit volume, and $e_{\rm int, 0}$ is the internal energy in the initial state. Indeed, in rare cases the prescriptions above can cause the kinetic energy to exceed the total energy. For example, consider a region where the mass inside the deposition zone is large enough that we are in the \texttt{MC} regime, but where there is a particular cell with very low density, such that the change in its kinetic energy density, $\Delta e_K = (p+\Delta p)^2 / [2 (\rho + \Delta\rho)] - p^2 / 2\rho$, is larger than the change in its total energy density. To ensure that the states we produce after deposition are valid, we apply a limiter to rescale the momentum update vector to enforce consistency. The state is consistent if
\begin{equation}\label{eq:lambda}
  C_V (\rho + \Delta \rho) T + \frac{|\bp + \lambda \Delta\bp|^2}{2 (\rho + \Delta \rho)} \le e + \Delta e,
\end{equation}
where $C_V$ is the gas specific heat capacity, and we determine the value of the limiter $\lambda$ by solving the resulting equation for $\lambda$ and choosing the largest root in the range $[0,1]$. The internal energy is then obtained by subtracting the kinetic energy from the total energy. In cases where the maximum solution for $\lambda$ equals 1, this procedure effectively heats up the gas. This can be necessary, for example, in regions between two SN explosions, where the momentum vectors from two depositions cancel each other and thus $|\Delta \bp|$ is close to zero; in this case our prescription correctly deposits the SN energy as heat.

\paragraph{Comparison to the original TIGRESS model.}
\label{para:tigress_comp}
Compared to the standard TIGRESS prescription outlined in \citet{Kim2017}, the primary changes to our approach are that we do not homogenise conditions inside the supernova deposition kernel, and we apply limiting only to the collective action of all supernovae rather than one supernova at a time. These changes lead to an algorithm that does not depend on the order in which supernova injection is processed, and that can cope with multiple supernovae depositing material in a single cell even if those particles reside on different parts of a decomposed domain -- both crucial features for the algorithm to operate smoothly on GPU. An added benefit of our approach even on CPU is that the TIGRESS prescription, which involves homogenising the material inside the supernova deposition zone prior to adding energy or momentum, produces a result that depends on the order in which particles are processed. This means either that one must adopt an arbitrary rule for which particle is processed first, or that the results are non-repeatable because the particles are processed in whatever order they happen to be stored in memory, which may vary depending on details such as the number of MPI ranks used in the simulation. Our prescription does not depend on arbitrary ordering choices, avoiding these undesirable features.

Another difference from the TIGRESS prescription is that we employ a fixed kernel size for all prescriptions, whereas TIGRESS varies the kernel size from $R_{\rm snr,min}=3\Delta x$ to a model-dependent $ R_{\rm snr,max} $ (typically 128 pc). TIGRESS adopts this variable approach to mitigate overcooling issues when the Sedov-Taylor phase is unresolved. In contrast, we observe no such overcooling issue in our code, even with a fixed $ R_{\rm snr} $ in the unresolved regime (see the test in \autoref{ssec:sn}). We attribute this to our use of a more accurate solver for temperature evolution: unlike TIGRESS, which relies on a single backward Euler step, we implement an ODE solver with adaptive time stepping to compute precise temperature updates. This enhancement avoids overcooling without requiring more than three ghost cells, thereby eliminating the need for additional communication beyond standard ghost-cell exchanges -- a critical advantage for efficient GPU implementation.

\subsubsection{Formation of supernova feedback particles}
\label{sec:sto}

Our method of depositing supernovae on the grid is independent of our choice for how to generate the particles that eventually produce those supernovae. However, for completeness here we also describe our default particle creation prescription in \quokka. Since our target resolution regime is one where individual massive stars can be resolved but the process of fragmentation that determines the initial mass function (IMF) cannot (typical cell mass at maximum resolution $\sim 1-10^2$ M$_\odot$), we implement a stochastic stellar population model to emulate the formation and evolution of individual massive stars. In this model, the star formation rate is controlled through two parameters: efficiency per free-fall time, $\epsilon_{\rm ff}$, and the fraction of gas cell mass that is converted into stars when star formation does occur, $\epsilon_*$. The first of these is a physical parameter describing an underlying theory for star formation on unresolved scales, which we set to $\epsilon_\mathrm{ff} = 0.01$ based on high-resolution observations of individual star-forming clouds \citep[e.g.,][]{Krumholz19a, Pokhrel21a, Hu22a}, while the latter is a numerical parameter that we set to $\epsilon_* = 0.5$, a choice that balances efficiency (favouring larger $\epsilon_*$, so that we avoid having to spawn stars repeatedly from the same cell) against stability (favouring smaller $\epsilon_*$, so that we perturb the hydrodynamic state minimally).

Given these parameters, our procedure is as follows: first, during each time step of size $\Delta t$, we identify cells on the finest adaptive mesh level where the density is above the \citet{Truelove1997} threshold given by \autoref{eq:truelove}. For this purpose we take $J=0.5$ to be conservative; this leaves us vulnerable to artificial fragmentation, but since in this regime we are prescribing the mass distribution of stars rather than attempting to calculate it directly, this is not a major concern. However, we do not place a star particle in every cell above the threshold density, since doing so would be computationally expensive. Instead, following the common approach in cosmological galaxy formation simulations, we treat star formation stochastically by assigning a probability $P$ of forming stars in any eligible cell. To determine $P$, note that the expected stellar mass formed from a cell of gas mass $M_\mathrm{cell}$
\begin{equation}
    \langle M_* \rangle = P \epsilon_* M_{\rm cell}\,.
\end{equation}
We then demand that $\langle M_* \rangle$ match the value predicted by the star formation efficiency per free-fall time $\epsilon_{\rm ff}$, which requires
\begin{equation}
    \langle M_* \rangle = \epsilon_{\rm ff} M_{\rm cell} \frac{\Delta t}{t_{\rm ff}}\,,
    \label{eq:mstar}
\end{equation}
where $t_{\rm ff} = \sqrt{3\pi/32 G \rho}$. Inverting the above two equations, the probability that a star that is eligible to do so forms a star during a time step is
\begin{equation}
    P = \min\left[\left(\frac{\epsilon_{\rm ff}}{\epsilon_*}\right) \left(\frac{\Delta t}{t_{\rm ff}}\right),1\right].
\end{equation}
In each time step we therefore randomly assign each cell that is eligible to form star to be star-forming or non-star-forming with probability $P$.\footnote{We note here that \autoref{eq:mstar} implicitly assumes that $\epsilon_{\rm ff}$ describes the star formation rate averaged over $\Delta t$. In principle one could also calculate the mass of stars formed by treating $\epsilon_\mathrm{ff}$ as describing an instantaneous star formation rate, and integrating the ordinary differential equation describing the gas mass, $dM_\mathrm{cell}/dt = -\epsilon_\mathrm{ff} M_\mathrm{cell}/t_\mathrm{ff}$, for a time $\Delta t$; one can then set $M_* = M_\mathrm{cell}(0) - M_\mathrm{cell}(\Delta t)$. For fixed $t_\mathrm{ff}$, this would yield $M_* = M_\mathrm{cell} (1 - e^{-\epsilon_\mathrm{ff} \Delta t/t_\mathrm{ff}})$, corresponding to $P = 1 - e^{-\epsilon_\mathrm{ff} \Delta t/t_\mathrm{ff}}$, the expression often adopted in cosmological star formation prescriptions. This choice and \autoref{eq:mstar} obviously yield very similar results if $\Delta t\ll t_{\rm ff}/\epsilon_\mathrm{ff}$, but the difference is non-negligible for larger time steps. That said, there is no good reason to assume that $t_\mathrm{ff}$ remains constant over such larger time steps, so there is no reason to believe that the exponential form is more accurate. In practice, however, the difference between the exponential form of $\langle M_*\rangle$ and \autoref{eq:mstar} is unimportant, both because the CFL condition almost always guarantees that $\Delta t\lesssim t_\mathrm{ff} \ll t_\mathrm{ff}/\epsilon_\mathrm{ff}$, and because if $\Delta t \gg t_\mathrm{ff}$ then we should not expect to obtain precisely the correct star formation rate from \textit{any} prescription based on an operator-split treatment of star formation and hydrodynamics.}

After identifying a cell for star formation, we spawn a stellar population drawn from a \citet{Chabrier05a} IMF, following the ``Poisson'' sampling strategy implemented by the \textsc{slug} stochastic stellar populations code \citep{Krumholz15b}. In our implementation, every stellar population comprises one particle representing the low-mass end of the IMF and a random number of individual high-mass stars; we draw the boundary between ``low-mass'' and ``high-mass'' at 9 M$_\odot$ because this is the minimum initial stellar mass (at Solar metallicity) predicted to produce a SNe in the \citet{Sukhbold16a} tables on which we rely. We draw the number of high-mass stars, $N_{\rm high}$, from a Poisson distribution with an expectation value $f_{*,\mathrm{high}} \epsilon_* M_\mathrm{cell} / \langle m_{*,\mathrm{high}}\rangle$, where $f_{\rm{*,high}} = 0.20055$ is the fraction of stellar mass contained in stars with masses $>9$ M$_\odot$, and $\langle m_*\rangle = 19.39$ M$_\odot$ is the mean mass per star for stars with masses $>9$ M$_\odot$; the numerical values given here are for our chosen IMF, but can be adjusted for different IMFs as desired. We then draw the masses of individual high-mass star particles from the high-mass end of the IMF, and set the mass of the low-mass star particle\footnote{We reiterate that this ``low-mass'' particle represents a collection of stars inhabiting the low-mass end of the IMF, unlike the ``high-mass'' particle(s) which represent a single high-mass ($>9$  M$_{\odot}$) star.} to $\epsilon_* M_\mathrm{cell}(1-f_{\rm{*,high}})$. 

It is important to note that in this implementation the mass of stars we produce will \textit{not} exactly equal the gas mass converted to stars on a cell-by-cell basis; instead the gas mass converted to stars is equal to the \textit{expectation value} of the stellar mass produced. Thus our star formation prescription guarantees mass conservation only in a statistical sense, not to machine precision. While this behaviour might seem undesirable, the alternative is worse: as discussed in \citet[see also \citealt{Haas10a}]{Krumholz15b}, there is no way to sample the distribution of stellar masses produced in events of finite total mass that simultaneously respects exact mass conservation within each event (i.e., that guarantees that the mass of stars drawn is exactly equal to the mass assigned to that event) and that also produces an IMF integrated over all star formation events that is independent of the mass assigned to each event. Nor is this a small effect: if the typical mass of a star-forming event is 100 M$_\odot$, then over many such events a prescription that requires the masses of stars formed in each event total exactly 100 M$_\odot$ could easily produce fewer than half as many supernovae as one would have obtained if the event size were 1000 or $10^4$ M$_\odot$. Thus if we were to enforce exact mass conservation, the result would be that simulations carried out at different resolutions (which determines the characteristic mass of cells that can form stars) would effectively have very different IMFs. Given a choice between this and reducing mass conservation from exact to statistical, we choose the latter as the lesser evil.

Once we have drawn high-mass star particles, we follow \citet{Andersson+21} and \citet{Steinwandel+23} by giving them initial velocities equal to the velocity of the cell which spawned them plus an additional component whose direction is random and whose magnitude is drawn from a power-law distribution $\propto v^{-1.8}$ with $v\in [3, 385]$ km s$^{-1}$. This distribution of velocities is motivated by simulations of dynamical ejections from young clusters \citep{Oh&Kroupa16}, and ensures creation of runaway stars with an incidence rate of $\sim 14\%$, matching empirical estimates -- see \citet{Andersson+21} and \citet{Steinwandel+23} for full details. To ensure no net momentum is added just by star particle creation, we set the velocity of the low-mass star particle such that, in the frame comoving with the cell that spawns the particles, its momentum is equal and opposite to the total high-mass momentum. 

\section{Tests}\label{sec:test}

We now present a series of tests of our implementations of sink and supernova feedback particles.
While the isothermal sphere test (\autoref{ssec:iso}), the Bondi/Bondi-Hoyle accretion tests (\autoref{ssec:bondi}, \autoref{ssec:bh}), and single SN feedback test (\autoref{sec:sn}) are not new, they demonstrate the correctness of multi-rank communication, as in all the tests we use multiple MPI ranks and place the particle at block boundaries.

\subsection{Self-similar collapse of an isothermal sphere}\label{ssec:iso}

\begin{figure}
    \centering
    \includegraphics[]{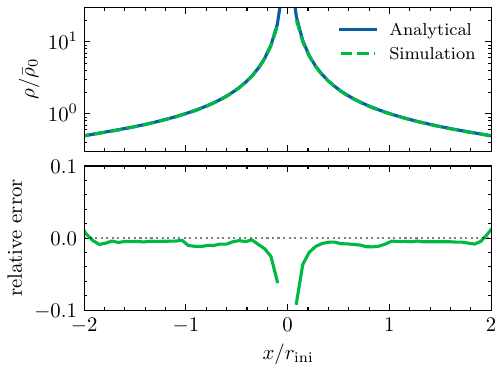}
    \includegraphics[]{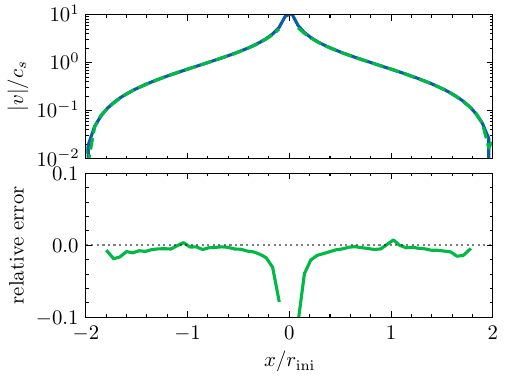}
    \includegraphics[]{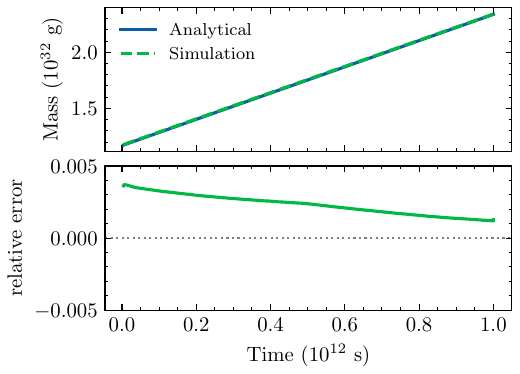}
    \caption{Isothermal sphere test. The top two panels show the mean density and radial velocity as a function of radius in the simulation (dashed green line) compared to the analytic solution (blue solid line). The lower panel shows the sink particle mass versus time, again comparing the simulation and exact analytic solutions.}
    \label{fig:iso}
\end{figure}

\edit{Our first test of sink accretion is the classic self-similar collapse of a singular isotehrmal sphere, which represents the inside-out gravitational collapse of a cold, nearly isothermal molecular cloud core forming a central protostar and infalling envelope \citep{Shu1977}. In this problem, the initial state is a spherically symmetric, isothermal gas cloud in unstable hydrostatic equilibrium, and the subsequent evolution describes \citeauthor{Shu1977}'s famous ``expansion-wave'' solution: the outer part retains a static singular isothermal equilibrium profile before the expansion wave arrives, and the inner part is free-falling toward the center, with the entire structure evolving self-similarly.}

% Our first test is to confirm our algorithms yield the correct accretion rate compared to the family of self-similar analytic solutions to accretion problems analysed by \cite{Shu1977}. 
\edit{The density profile for the outer, hydrostatic part of a generalized singular isothermal sphere is 
\begin{equation}
    \rho(r)=\frac{A c_s^2}{4 \pi G r^2}.
\end{equation}
where $c_s$ is the isothermal sound speed and $A$ is a dimensionless parameter that measure the degree of over-density relative to the marginally stable equilibrium.
The special case $A=2$ corresponds to an isothermal sphere that is in unstable hydrostatic equilibrium everywhere outside the expansion wave.}

We consider a sphere of isothermal gas with sound speed $c_s = 0.2~{\rm km/s}$. To avoid the singular initial configuration of the \citeauthor{Shu1977} solution, we do not start from the $t=0$ state with a formal $\rho \propto r^{-2}$ divergence at the origin. Instead, we initialize our density and velocity profile to the analytic solution at a finite time $t_0 = 10^{12}$ s, by which point a central point mass has already formed an the expansion wave has propagated outward\edit{; the choice of initial time is essentially arbitrary, since the analytic solution is self-similar -- the structure is the same at all times $>0$, and the choice of $t_0$ simply serves to set the initial position of the expansion wave}. For $x \in [0, 1]$, where $x = r / c_s t$ is the similarity variable, we use the asymptotic solution of the expansion-wave collapse problem, taking the case $A = 2.0001$ as an approximation to the $A = 2^+$ limit; for $x > 1$, we adopt the isothermal solution $v = 0, \ \alpha = 2/x^2$. We run the simulation in cgs units; to convert the dimensionless variables in the initial condition to dimensional quantities, we use the following definitions of the unit length, mass, velocity, and density:
\begin{gather}
    u_l = c_s t_0 = 2 \times 10^{16} \ {\rm cm} \\
    u_m = c_s^3 t_0 / G = 1.198627571 \times 10^{32} \ {\rm g} \\
    u_v = c_s = 0.2 \ {\rm km/s} \\
    u_{\rho} = \frac{1}{4 \pi G t_0^2} = 1.192296893 \times 10^{-18} \ {\rm g~cm^{-3}}.
\end{gather}
Initially, we place a sink particle with mass $m_0 = 0.975 ~u_m = 1.168661882 \times 10^{32}$ g at the domain centre. The computational domain is a box with size $8 u_l$, and we use a base grid of $128^3$ cells with two refinement levels to achieve a maximum resolution $\Delta x = u_l / 64$ within $2u_l$ of the origin.

In the top two panels of \autoref{fig:iso}, we plot the density and velocity profiles at $t = 10^{12}$ s, enough time for the expansion wave to have expanded in radius by a factor of two; we plot the sink particle mass versus time in the lower panel. The numerical results agree with the analytic solution extremely well: the relative error in both density and velocity remains within $2\%$ across radii inside the expansion wave front, with larger deviations only in the innermost region where spatial resolution is poor. The temporal evolution of the sink particle mass is in excellent agreement with theoretical expectations.

\subsection{Bondi accretion}
\label{ssec:bondi}

\begin{figure*}
    \centering
    \includegraphics[width=0.9\linewidth]{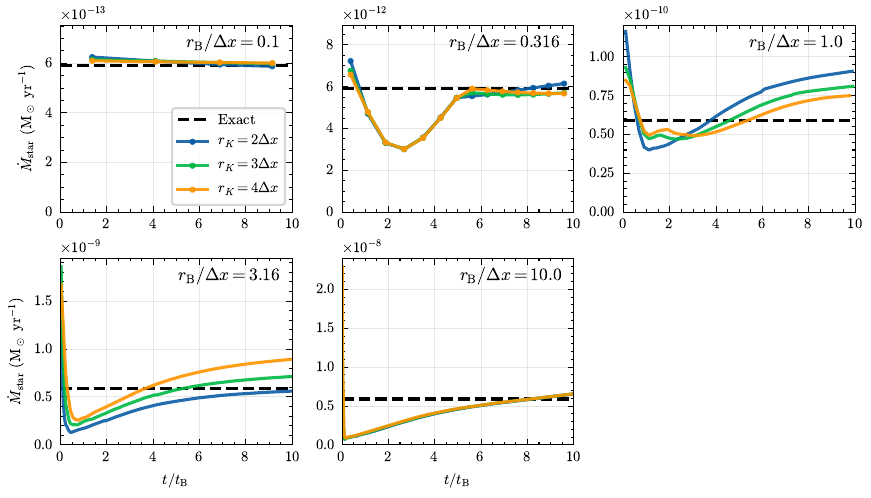}
    \caption{Comparison of simulated accretion rates over time with theoretical values for the Bondi accretion tests presented in \autoref{ssec:bondi}. The ratios of the Bondi radius to spatial resolution are, from left to right and top to bottom, 0.1, 0.32, 1.0, 3.16, and 10. The dashed horizontal line denotes the theoretical accretion rate, and the solid lines show the accretion rate measured in the simulations. The horizontal axis gives time in units of the Bondi time.}
    \label{fig:sink_bondi}
\end{figure*}

Our second test of the sink particle algorithm is against the \citet{Bondi52a} solution for accretion onto a point mass by an infinite, uniform medium with negligible self-gravity. Since we have built the Bondi accretion formula into our method (\autoref{eq:mdot}), we expect it to reproduce the exact analytic accretion rate in the unresolved regime $r_{\rm B} \ll \Delta x$, where $r_\mathrm{B}$ is the Bondi radius given by \autoref{eq:rbh} with $v_\infty = 0$. In the opposite regime, $r_{\rm B} \gg \Delta x$, the accretion rate should depend primarily on the hydrodynamics and gravity solver, so most reasonable schemes for accretion by a point mass should yield an accurate solution, but the marginally resolved case ($r_{\rm B} \sim \Delta x$) is more challenging. To assess the performance of our scheme, we conduct a suite of simulations with largely identical initial conditions but a range of values for $r_{\rm B} / \Delta x$, following the setup of \cite{Krumholz2004}. We initialize a gaseous sphere of radius $r_{\rm sph} = 1.21 \times 10^{19}$ cm at a constant temperature of 10 K with mean molecular weight $\mu = 2.33 ~m_p$ with a sink particle of mass $m_s$ at its centre. The gas sphere is centred in a box of size $16 ~r_{\rm sph}$. The mesh comprises a coarse level of 128 cells, with level-1 refinement covering a box of size $8~r_{\rm sph}$, and level-2 refinement covering $4~r_{\rm sph}$. With $m_s = 1~M_{\odot}$, this configuration yields $\Delta x = r_{\rm B}$ at the finest level. We then vary $m_s$ so that $r_{\rm B} / \Delta x$ ranges from 0.1 to 10 in steps of 0.5 dex. In all cases we initialize the density and velocity profiles of the sphere to the analytic Bondi solution, which is given implicitly by the solutions to the system of coupled ordinary differential equations  
\begin{align}
x^2 \alpha v  & = \lambda\\
\frac{v^2}{2} + \ln(\alpha) - \frac{1}{x} & = 0,
\end{align}
where $x \equiv r/r_{{\rm B}}, v \equiv |u|/c_{\infty}, \alpha \equiv \rho / \rho_{\infty}$, and $\lambda$ is the mass accretion rate in units of the fiducial mass flux $\rho_{\infty} c_{\infty}$ through area $4 \pi r_B^2$,
\begin{equation}
\lambda \equiv \frac{\dot{M}}{ 4 \pi \rho_{\infty} (GM)^2 / c_{\infty}^3 }.
\end{equation}
We allow the simulation to evolve until the accretion rate reaches a steady state; the natural unit of time for this system is $t_{\rm B} = r_{\rm B}/c_s$, and the time to reach equilibrum is generally $\sim 5 ~t_\mathrm{B}$.

The results are summarized in \autoref{fig:sink_bondi}. We obtain high accuracy (error $< 2\%$) in the steady-state accretion rate when the Bondi radius $r_{B}$  is 10 times smaller than the cell spacing $\Delta x$. When $r_B$ and $\Delta x$ are comparable, the error increases to $\sim 20\%$. As argued by \cite{Krumholz2004}, the larger error for $r_{\rm B} \approx \Delta x$ is expected because the transition from subsonic to supersonic accretion flow is poorly resolved, leading to substantial errors in the density and velocity in cells near the sink particle. For $r_B = 10 \Delta x$ the error at $t = 10 t_{\rm B}$ drops to a few percent. 

We also use this test to explore kernel sizes from $r_K = 2 \Delta x$ to $4 \Delta x$ and find that the choice of $r_K$ makes only minor differences when $r_{\rm B}/\Delta x \sim 1$; in particular, at $r_{\rm B}/\Delta x = 1$, larger $r_{\rm K}$ yields slightly improved accuracy. However, for $r_B / \Delta x \gg 1$ or $r_B / \Delta x \ll 1$, the difference is negligible. This behaviour is consistent with the arguments above regarding the scheme's accuracy in the well-resolved and unresolved limits.

\subsection{Bondi-Hoyle accretion}
\label{ssec:bh}

\begin{figure}
    \includegraphics[width=0.84\columnwidth]{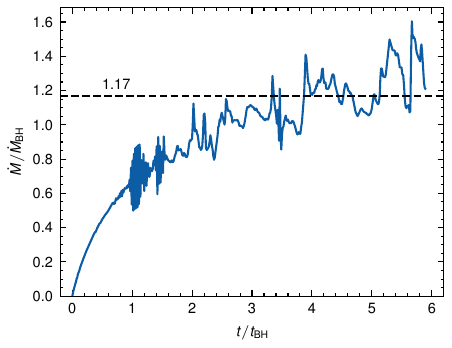}
    \includegraphics[width=0.95\columnwidth]{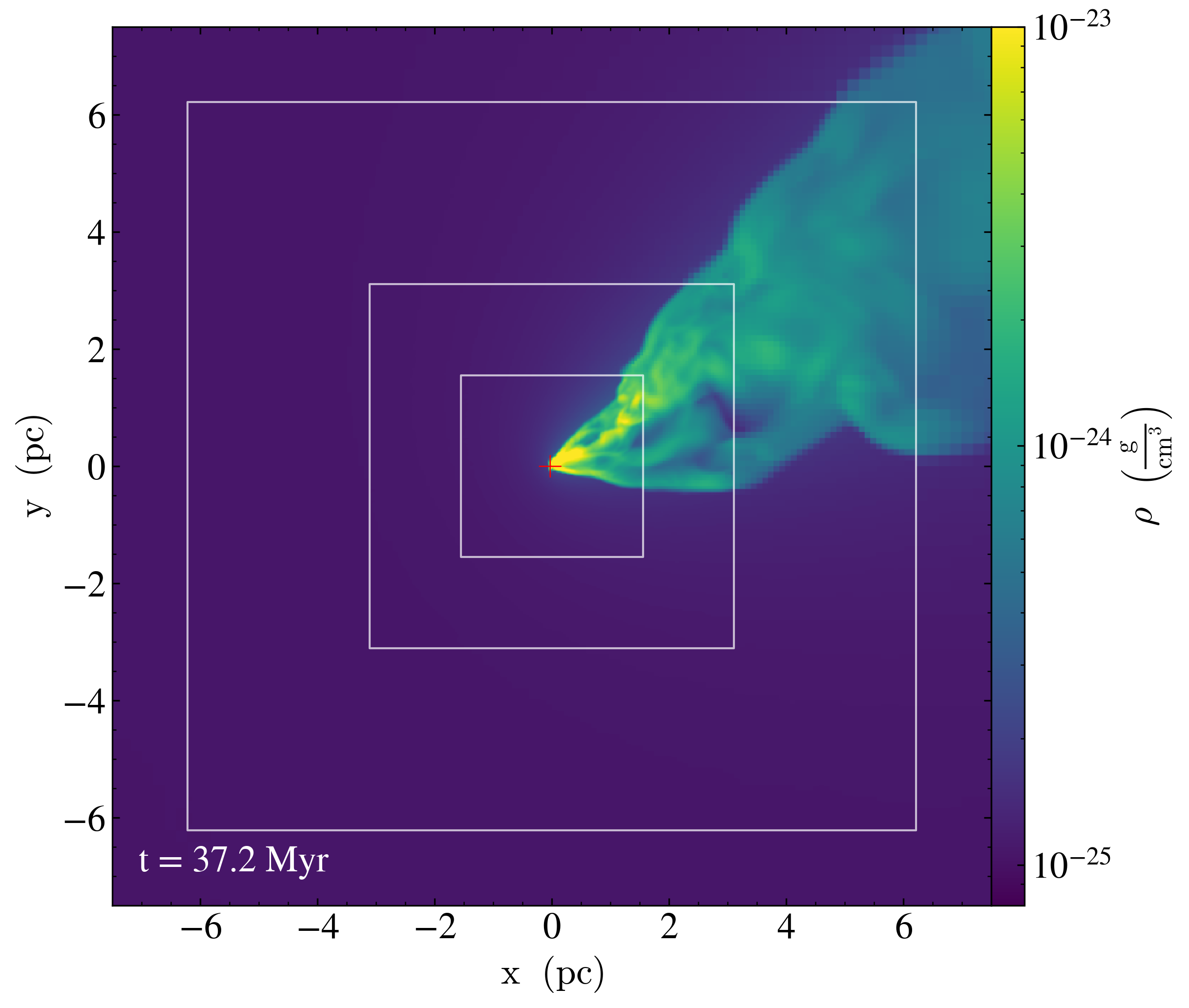}
    \caption{Top: Accretion rate versus time for the Bondi-Hoyle accretion test described in \autoref{ssec:bh}. The accretion rate is normalized to the Bondi-Hoyle accretion rate, and the time is normalized to Bondi-Hoyle time. Bottom: Snapshot of the density field in the $xy$-plane at $t \approx ~6 t_{\rm BH}$. White squares indicate the borders of the AMR levels.}
    \label{fig:bh}
\end{figure}

We evaluate our sink particle scheme's ability to model accretion onto a moving particle by simulating the classical Bondi-Hoyle accretion problem. For this test, we initialize the system in the regime  $r_{\rm BH} / \Delta x \gg 1$. We use the same grid hierarchy as in \autoref{ssec:bondi}, but reduce the domain size -- and thus the grid spacing -- by a factor of 5.
The gas has a uniform density $\rho = 10^{-25}~{\rm g ~cm^{-3}}$ and an initial velocity with Mach number $\mathcal{M} = 3$. The velocity lies in the $x-y$ plane, forming a $30^\circ$ angle with the $x$-axis. We place a $100~M_{\odot}$ sink particle at the domain center. Using \autoref{eq:rbh}, we obtain the Bondi-Hoyle radius $r_{\rm BH} = 0.1 ~r_{\rm B}$, which in turn yields $r_{\rm BH} / \Delta x = 100 \times 0.1 \times 5 = 50$ (see calculations in \autoref{ssec:bondi}).

\autoref{fig:bh} shows the accretion rate versus time. It takes several Bondi-Hoyle times, $t_\mathrm{BH} = r_{\mathrm{BH}} / v_\infty$, for the system to reach a quasi-steady value. As discussed by \cite{Krumholz2004}, this setup is in a regime where the interpolation formula is not particularly accurate. \cite{Ruffert1996} and \cite{Krumholz2004} performed similar simulations and found a steady-state accretion rate close to $\dot{M} \approx 1.17 \dot{M}_{\rm BH}$, where $\dot{M}_{\rm BH}$ is defined in \autoref{eq:mdot}. Our results agree well with these studies: the difference in the steady-state mean accretion rate is smaller than the intrinsic fluctuations. Both works also reported temporal variability in the accretion rate and flow morphology, and suggested Rayleigh-Taylor and Kelvin-Helmholtz instabilities as plausible causes. 

\subsection{Single SN feedback}\label{sec:sn}

\begin{figure}
    \centering
    \includegraphics[]{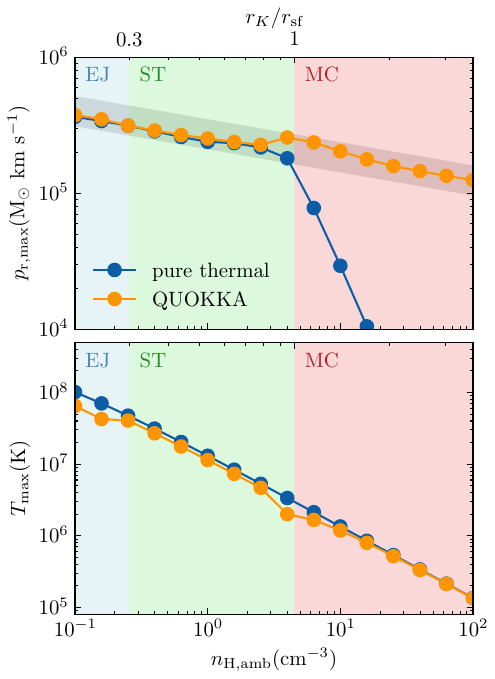}
    \caption{Tests of SN feedback prescriptions using simulations of radiative SNR evolution in a uniform medium with spatial resolution $\Delta x = 4$ pc, both using our fiducial prescription (orange) and a scheme that deposits thermal energy only (blue). The top panel shows the final radial momentum, while the bottom shows the maximum temperature attained during the simulation, both shown as functions of the ambient hydrogen number density. The grey stripe in the upper panel indicates the empirical $p_{\rm r,max} - n_{\rm H, amb}$ relation with $25\%$ tolerance. The upper axis shows the ratio of kernel radius ($=3\Delta x$) to the shell-formation radius, and the background shading indicates which case of our fiducial SN deposition scheme is used for each simulation given that ratio.}
    \label{fig:psnr}
\end{figure}

We assess the accuracy of our SN feedback implementation by simulating SNR evolution with radiative cooling in a uniform medium at fixed spatial resolution $\Delta x = 4$ pc in a domain of size 256 pc and no adaptivity, following \cite{Kim2017}. We vary the ambient density from $n_{\rm amb} = 0.1~{\rm cm}^{-3}$ to $100~{\rm cm}^{-3}$. We initialize each simulation with a SN particle that explodes at $t=0$, and we evolve the problem to the point where the total radial momentum in the domain, defined as $p_\mathrm{r} = \int \rho \mathbf{v} \cdot \hat{\mathbf{r}} \, dV$, reaches steady-state. \autoref{fig:psnr} presents the final radial momentum (top) and the maximum temperature achieved in any cell over the full simulation time (bottom) as a function of $n_{\rm amb}$; the corresponding ratios of $r_{K} = 3 \Delta x$ to shell-formation radius $r_{\rm sf} = 22.6~{\rm pc} (n_{\rm amb} / {\rm cm}^{-3})^{-0.42}$ are indicated on the top axis. Results for our \quokka{} prescription are shown in orange. For comparison, we also show the result of the ``pure thermal'' model in blue, which injects thermal energy at all resolutions.

In the \texttt{EJ} regime, where $\mathcal{R}_M \le 0.027$ (roughly $r_K / r_{\rm sf} \le 0.3 $), our fiducial model applies pure kinetic energy deposition. In this regime the ST stage is fully resolved, and \citet{Kim2015} find that both thermal and kinetic deposition recover the correct final radial momentum. Consistent with this expectation, our fiducial model coincides with the pure-thermal model in the top panel. The maximum temperature reached with kinetic deposition is only slightly lower than with pure-thermal deposition because in kinetic mode the gas is shock-heated shortly after SN ejection.

As the Sedov-Taylor stage becomes marginally resolved, $0.3 < r_K / r_{\rm sf} < 1$, the prescriptions diverge at the level up to a few tens of percent; in the unresolved regime ($r_K / r_{\rm sf} > 1$), the discrepancies are even larger. In these regimes our fiducial model (orange) remains within $25\%$ of the empirical $p_{\rm r, max} - n_{\rm H,amb}$ relation (shaded area), demonstrating the the scheme is robust. By contrast, the pure-thermal model deviates substantially from the empirical relation owing to overcooling, consistent with previous studies \citep[e.g.][]{Kim2015, Kim2017}. Interestingly, while still incorrect, our pure thermal calculations yield a curve of $p_{\rm r, max}$ versus $n_\mathrm{H,amb}$ that lies closer to the empirical relation than reported by \cite{Kim2017}, which we attribute to improvements in our cooling solver (see \autoref{para:tigress_comp}).

\subsection{Multiple SN feedback}

\begin{figure}
    \centering
    \includegraphics[scale=0.34]{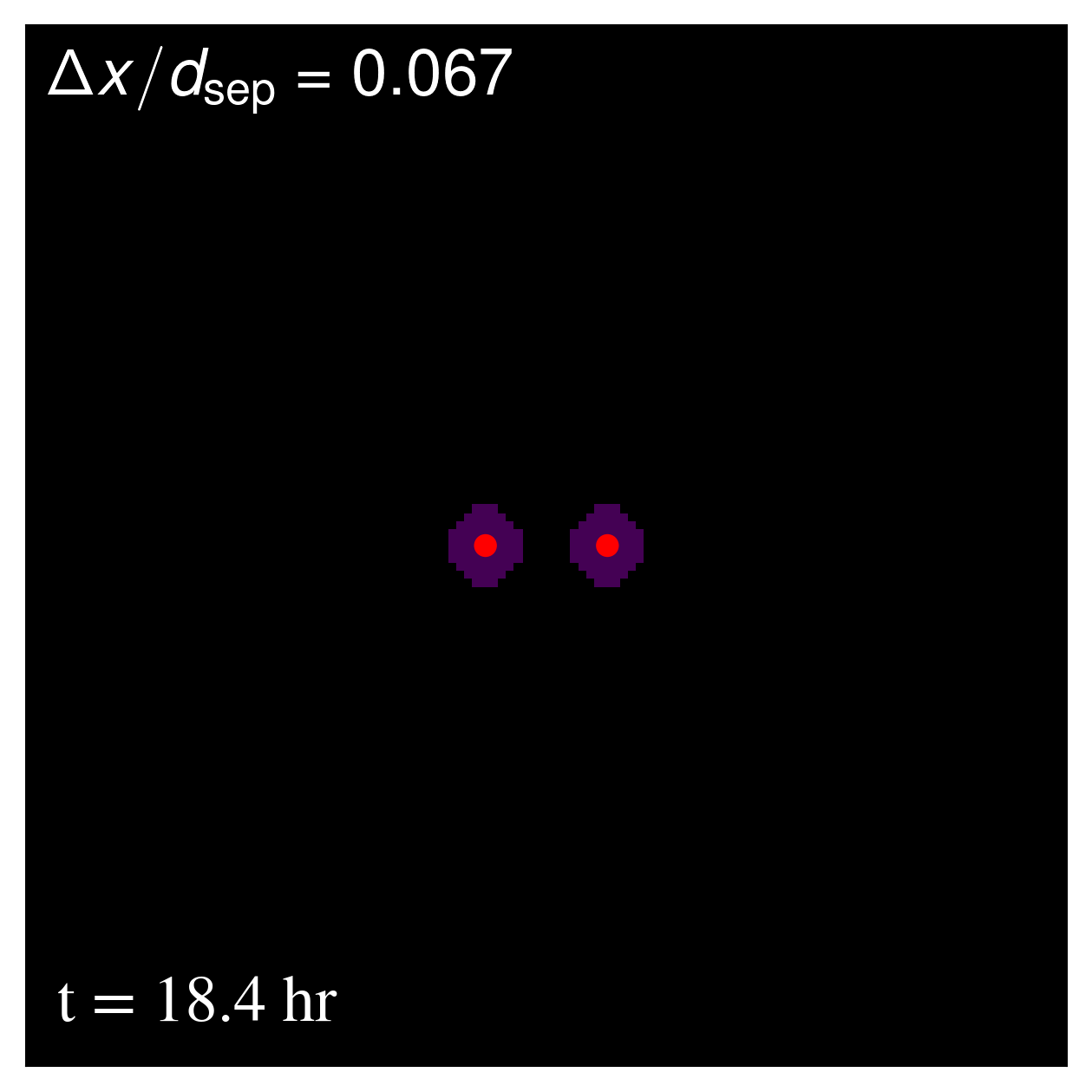}\includegraphics[scale=0.34]{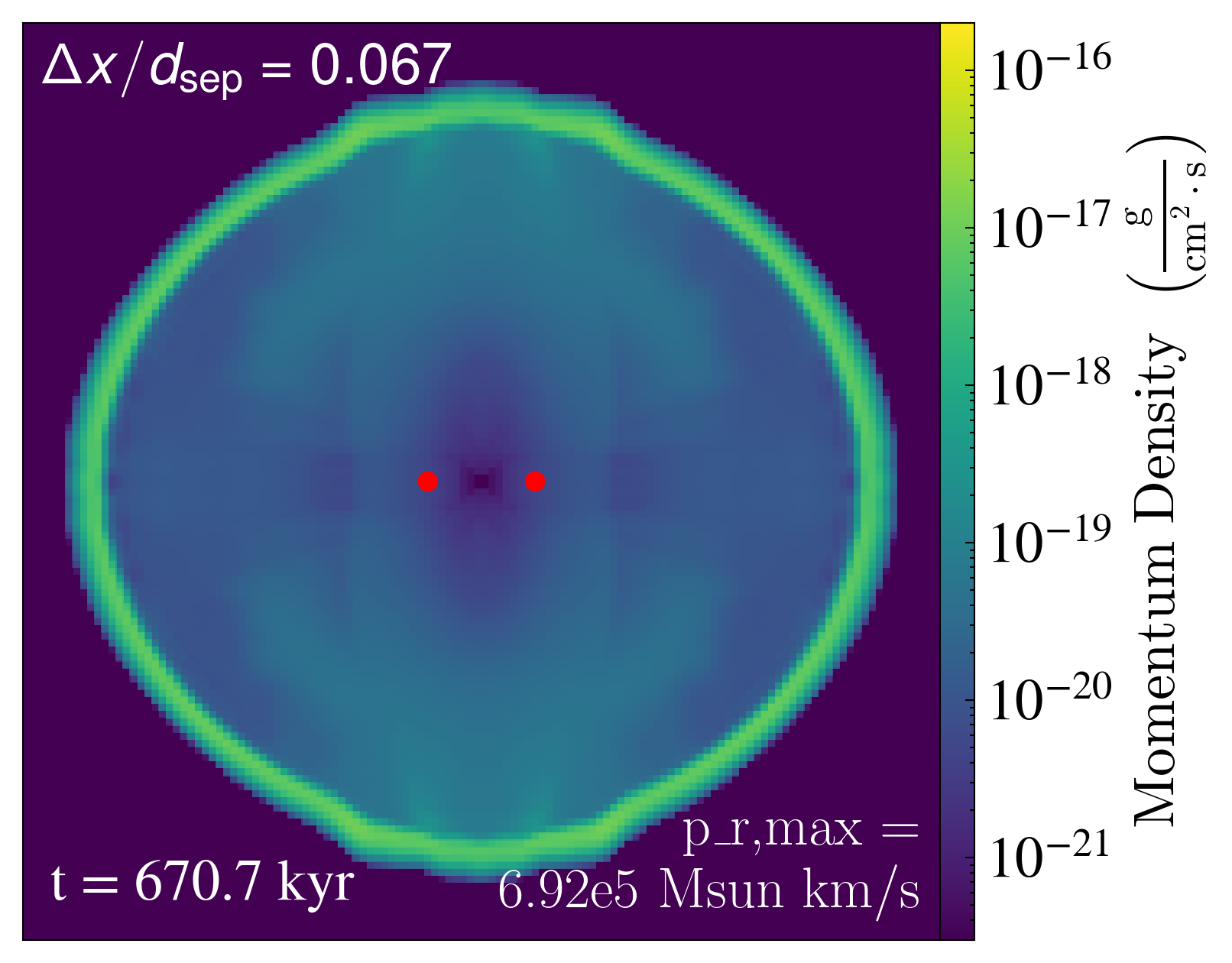}\\
    \includegraphics[scale=0.34]{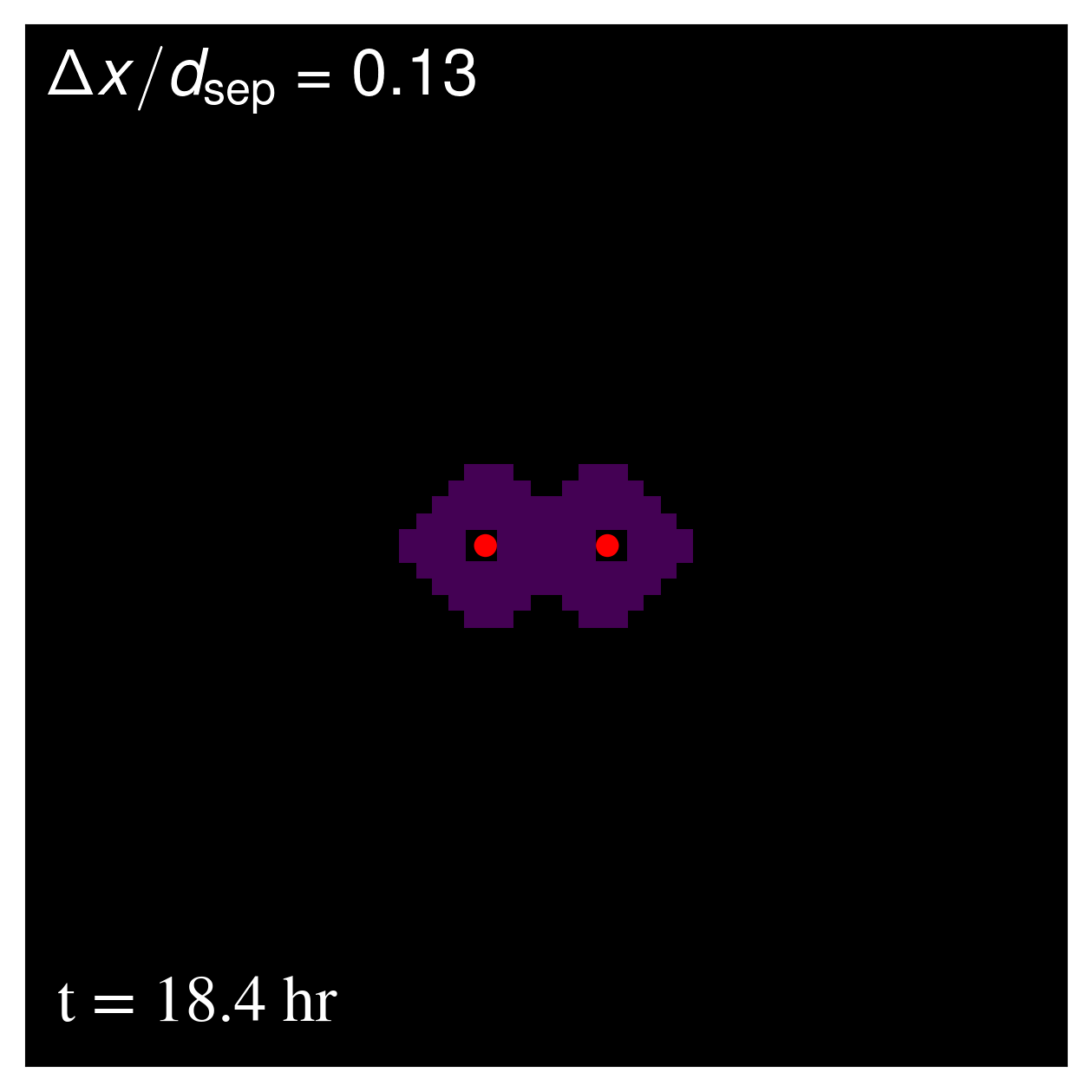}\includegraphics[scale=0.34]{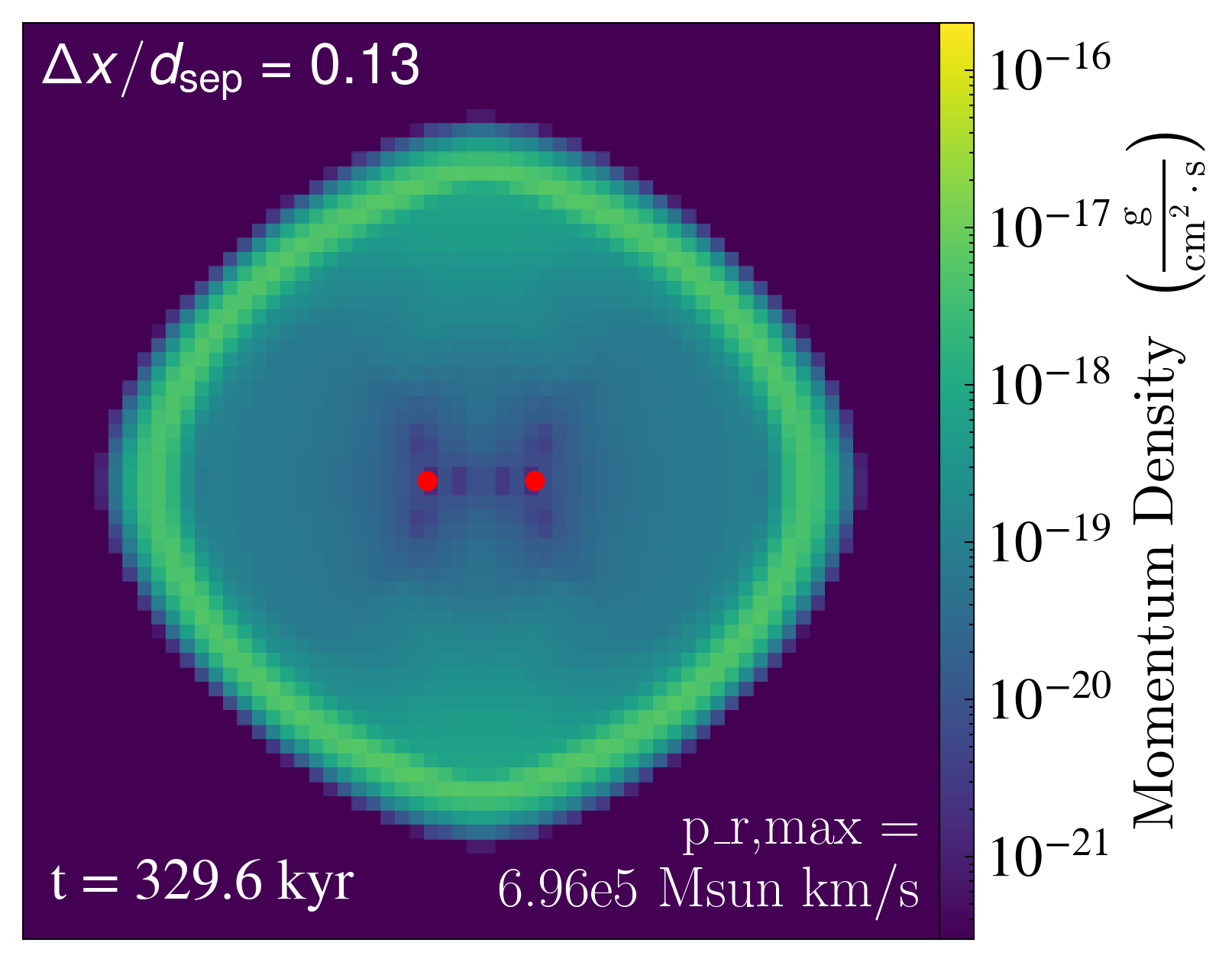}\\
    \includegraphics[scale=0.34]{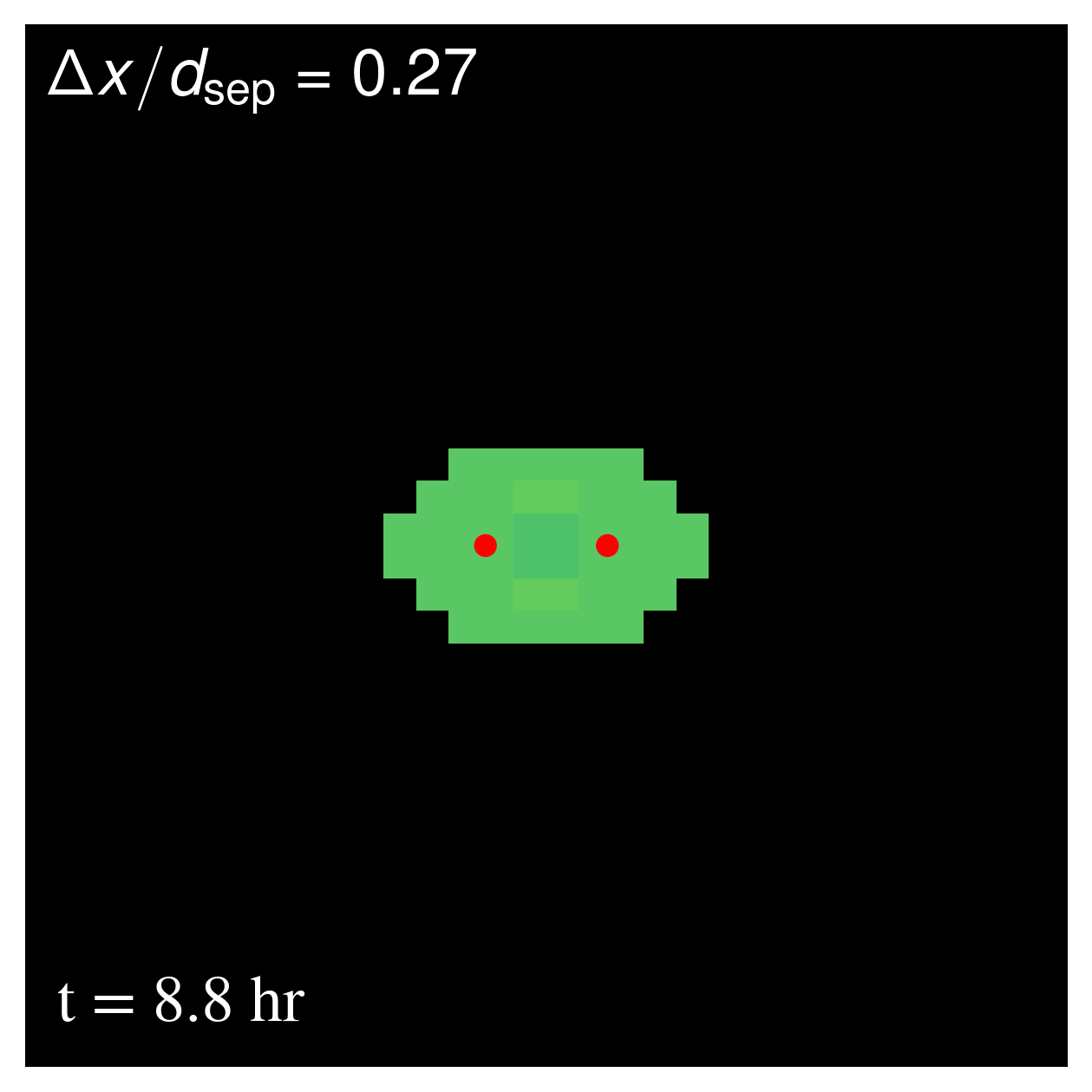}\includegraphics[scale=0.34]{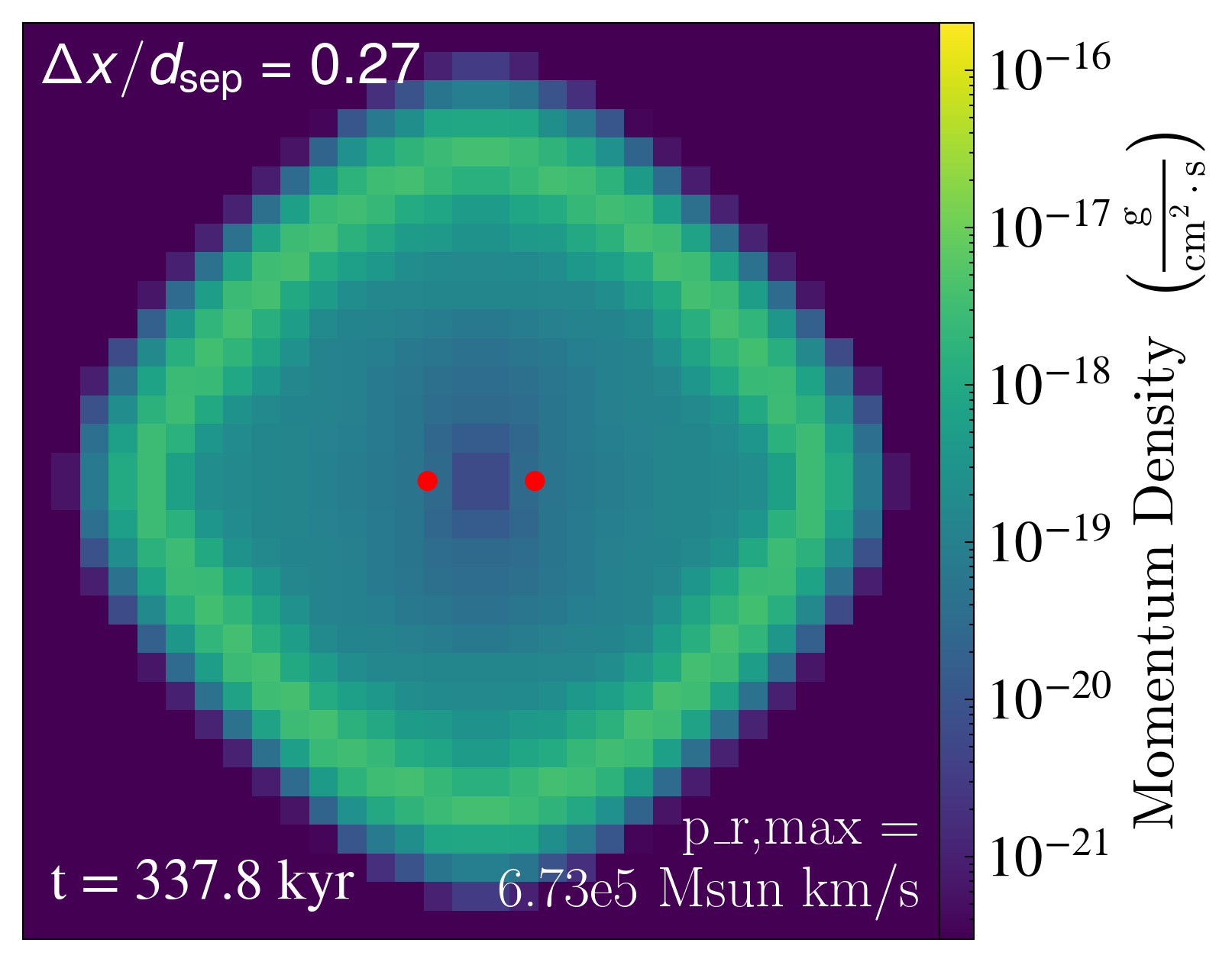}\\
    \includegraphics[scale=0.34]{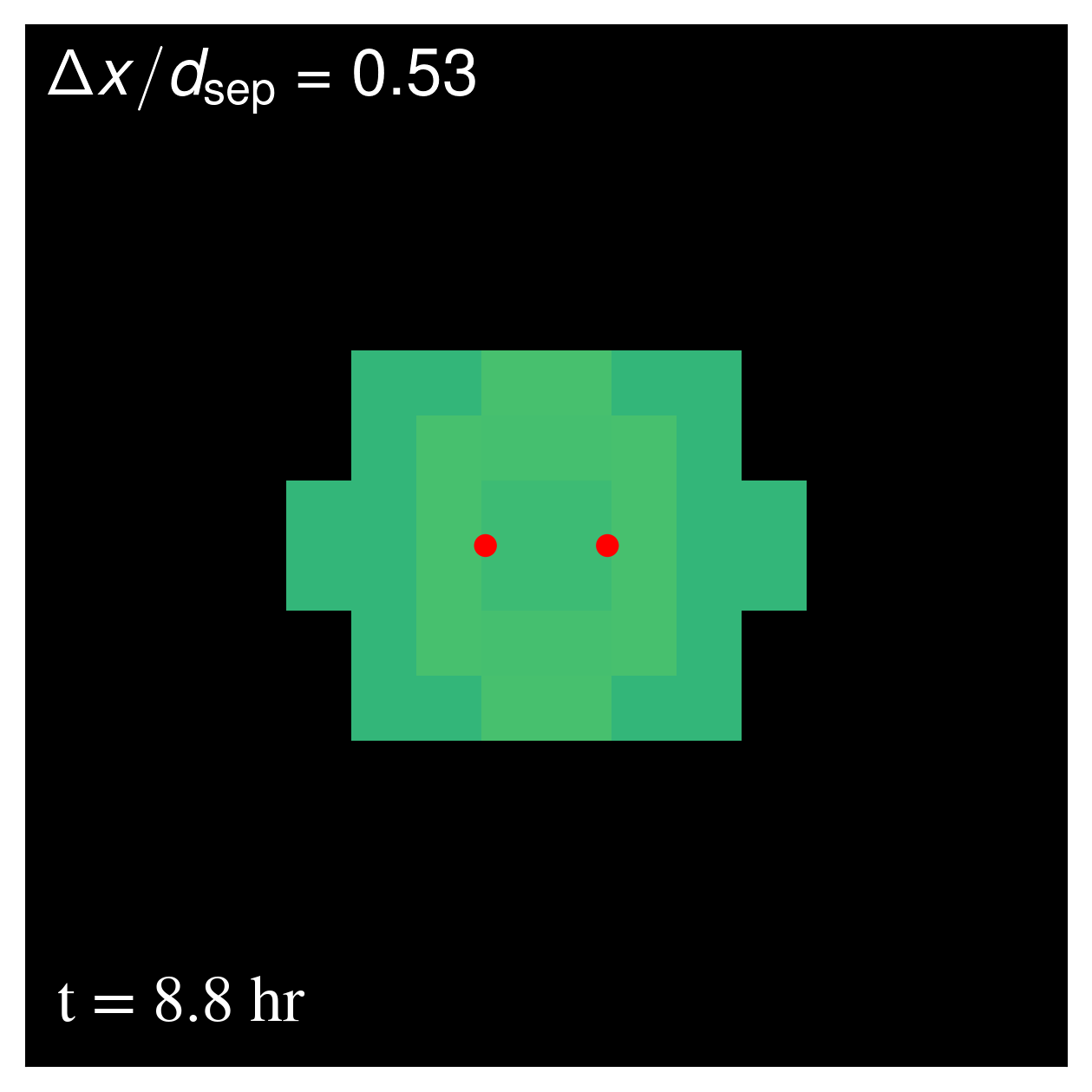}\includegraphics[scale=0.34]{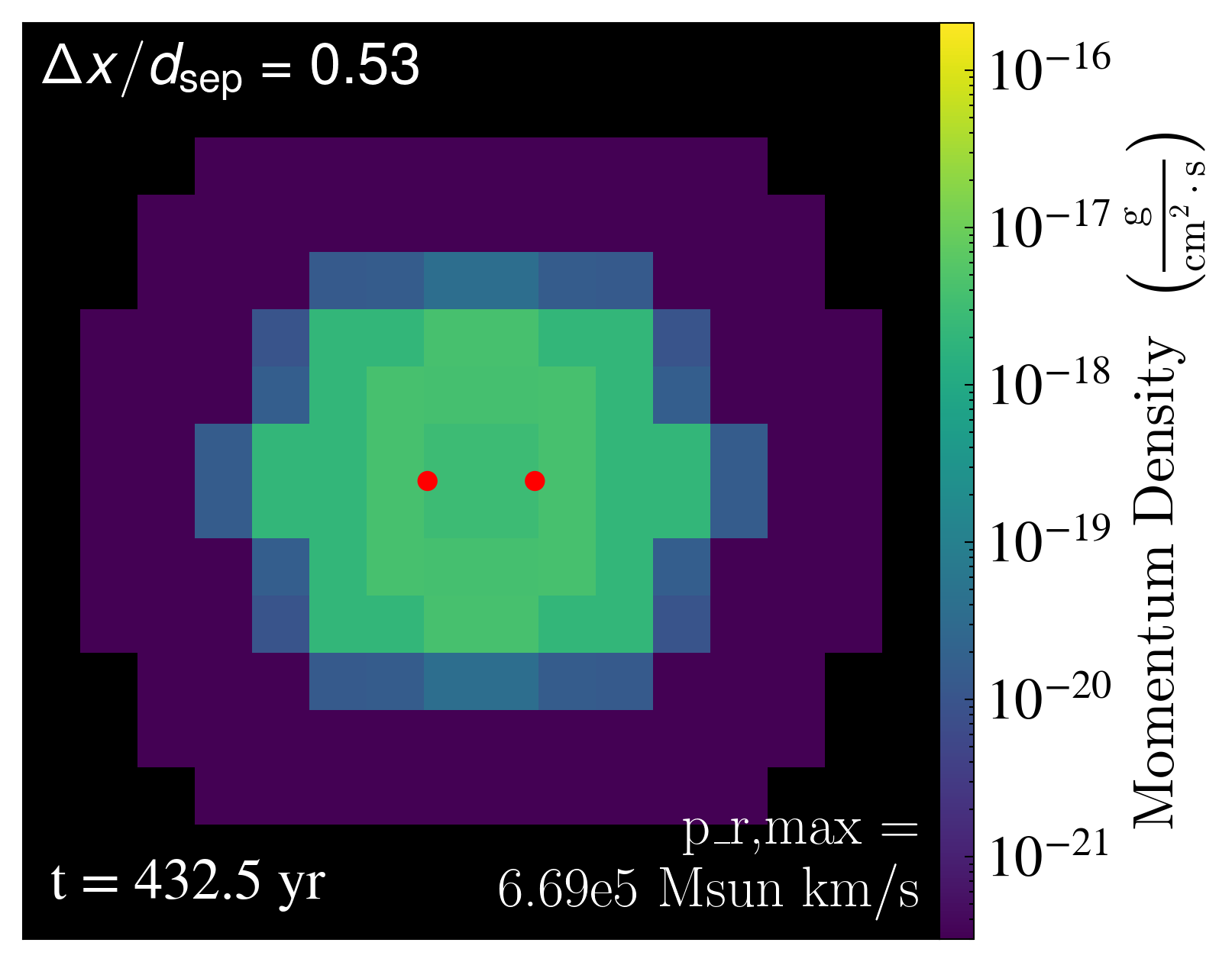}
    \caption{Slices showing radial momentum density in the double-SN test with $r_{\rm sf} / d_{\rm sep} \approx 2.0$ (corresponding to ambient density $n_\mathrm{H,amb} = 0.1$ cm$^{-3}$), so shell formation occurs only {\it after} the two SNRs begin to interact. Red circles mark the locations of the two SNe. The left column shows the configuration after the first time step, immediately after SN deposition, while the right column shows the state at the end of the simulation when the radial momentum has ceased increasing, at the time indicated.  Spatial resolutions are 2 pc, 4 pc, 8 pc, and 16 pc, from top to bottom. Our numerical method demonstrates converging results for the terminal radial momentum (see labels at bottom-right corner of the second column) as we vary the spatial resolution.
    }
    \label{fig:sn22}
\end{figure}

\begin{figure}
    \centering
    \includegraphics[scale=0.34]{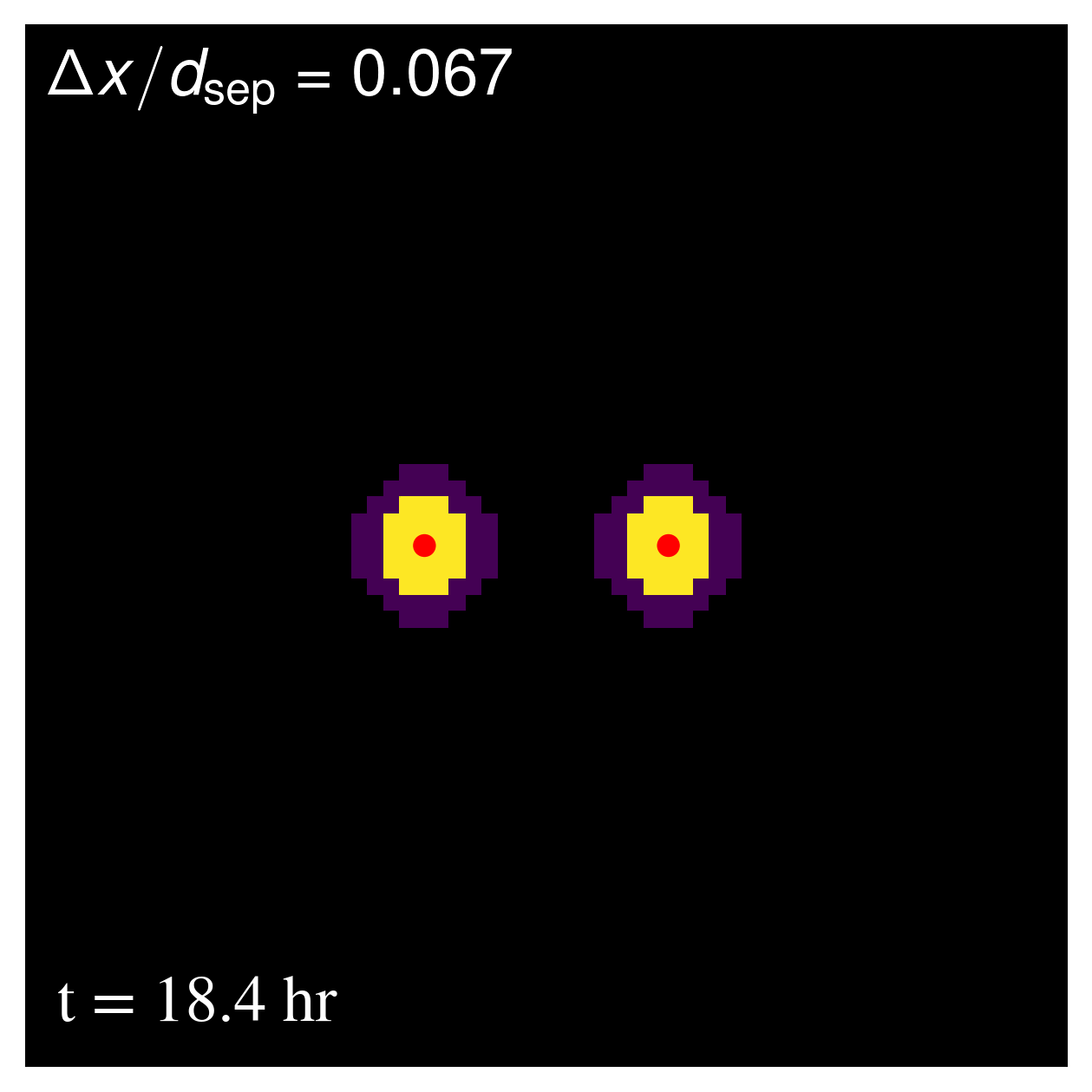}\includegraphics[scale=0.34]{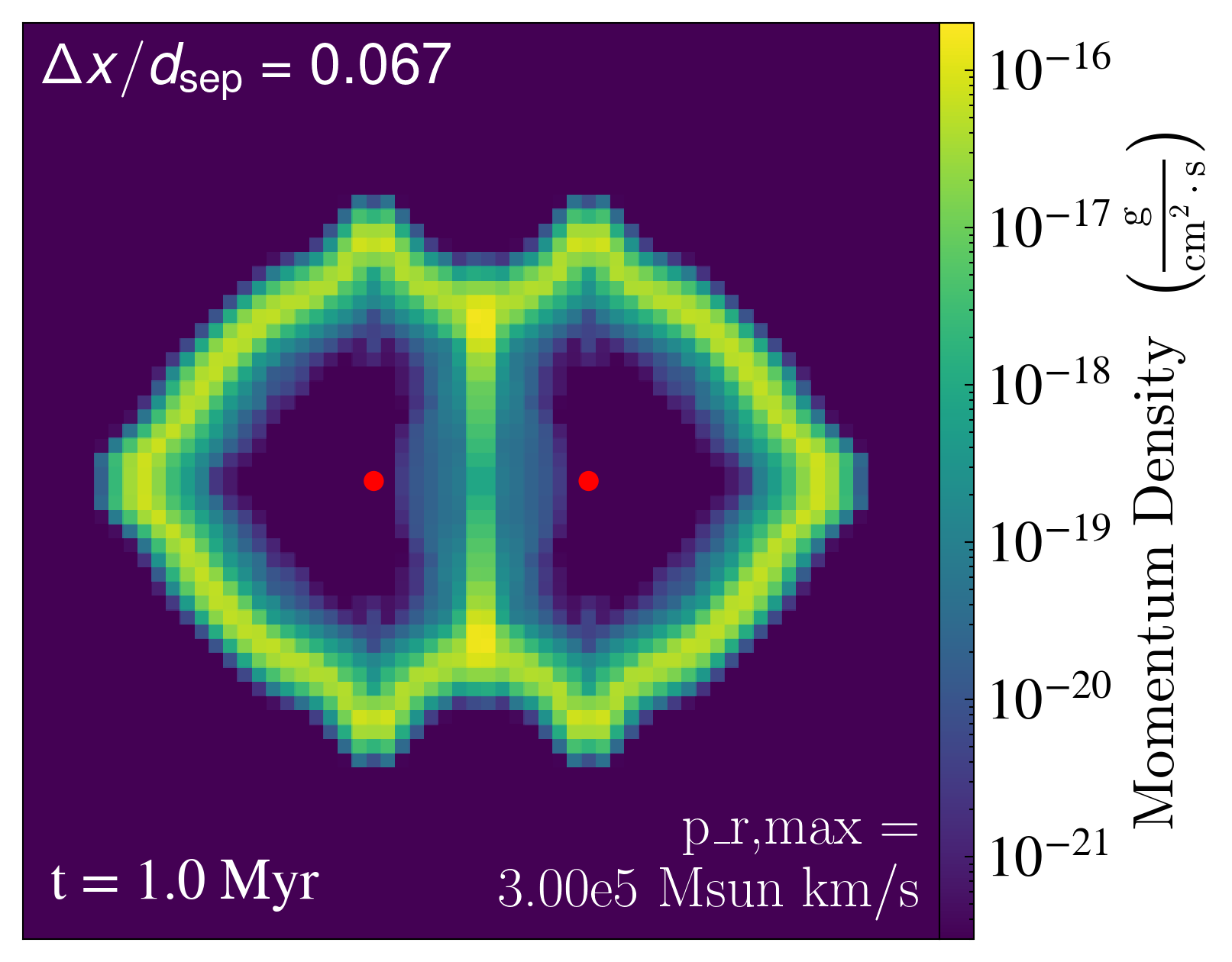}\\
    \includegraphics[scale=0.34]{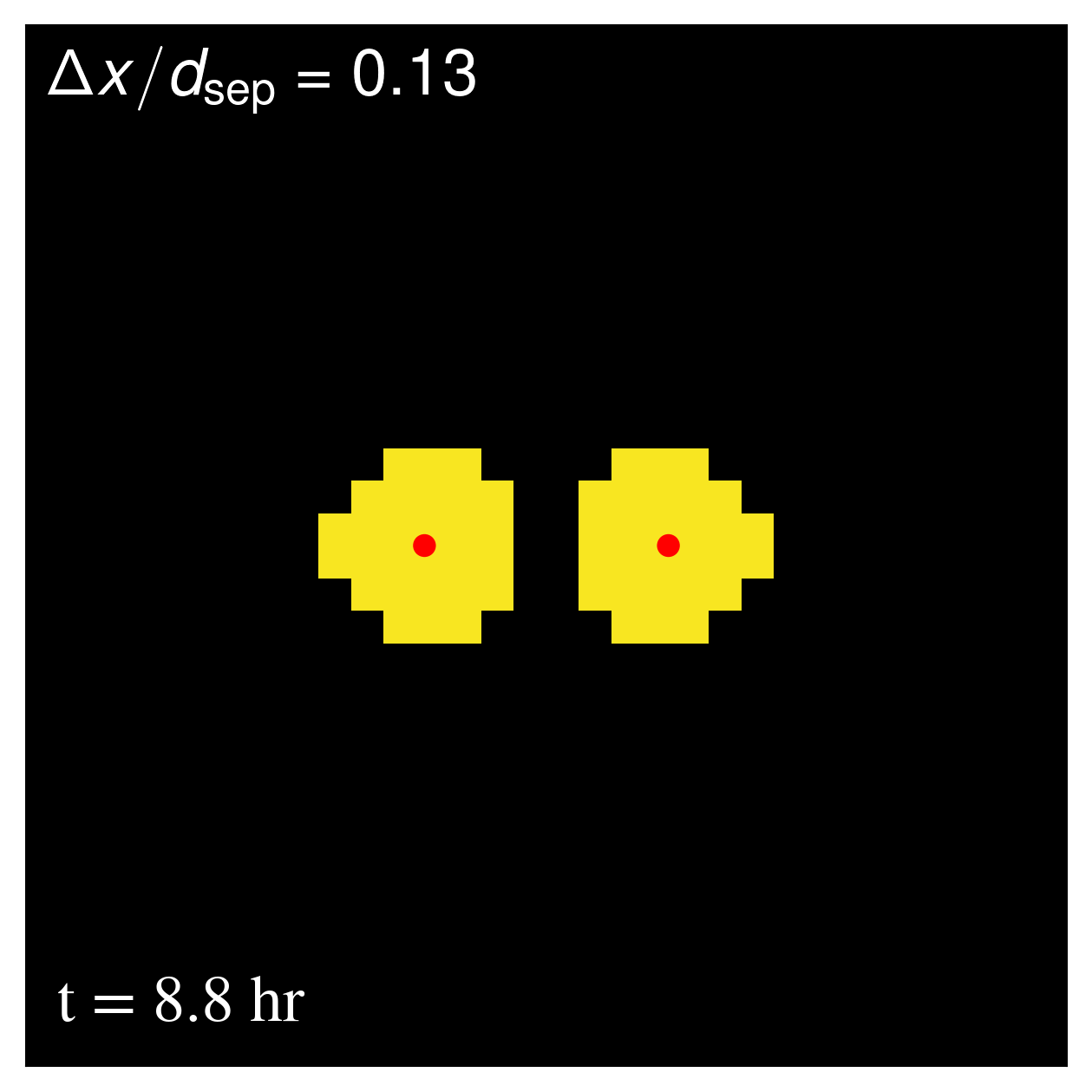}\includegraphics[scale=0.34]{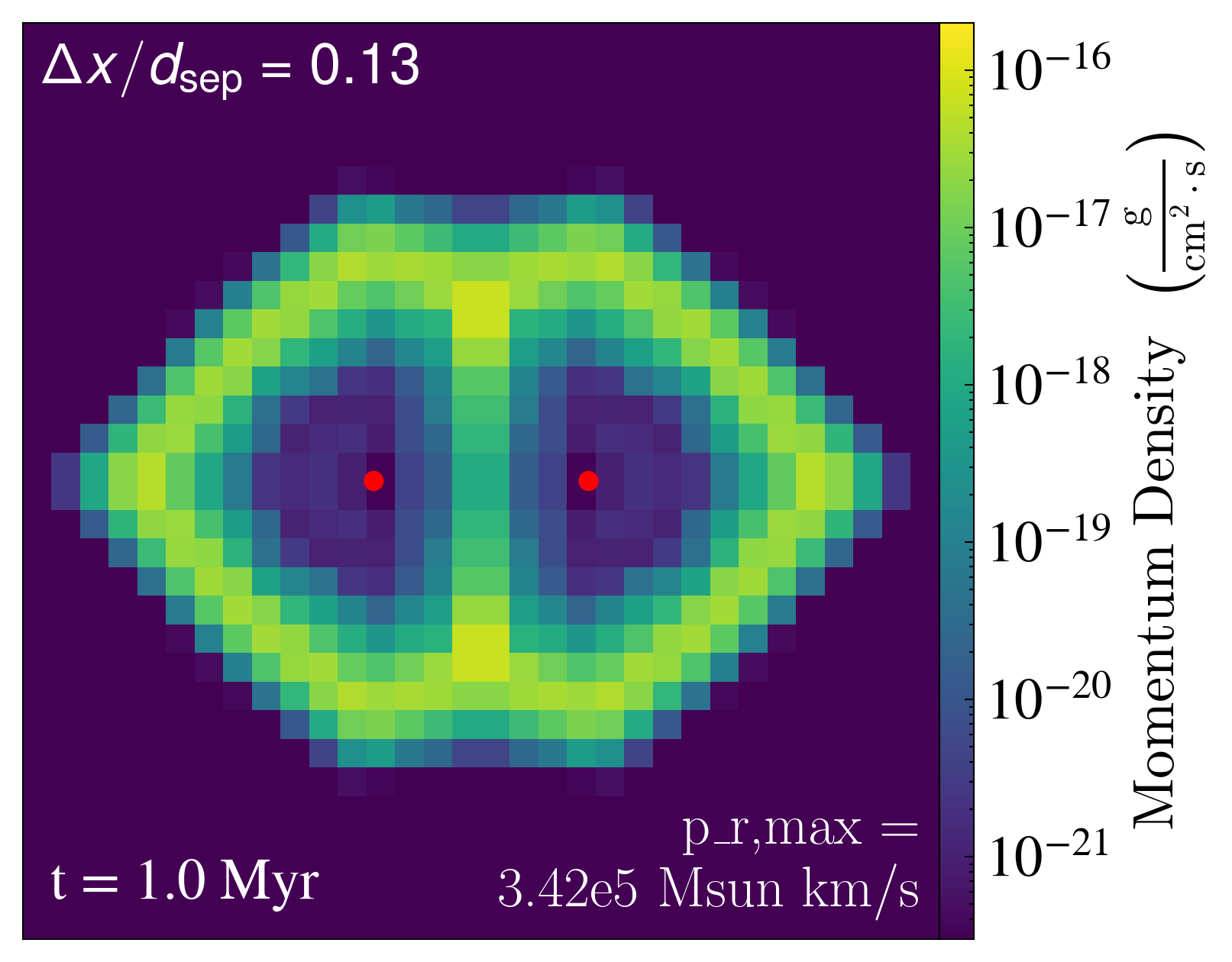}\\
    \includegraphics[scale=0.34]{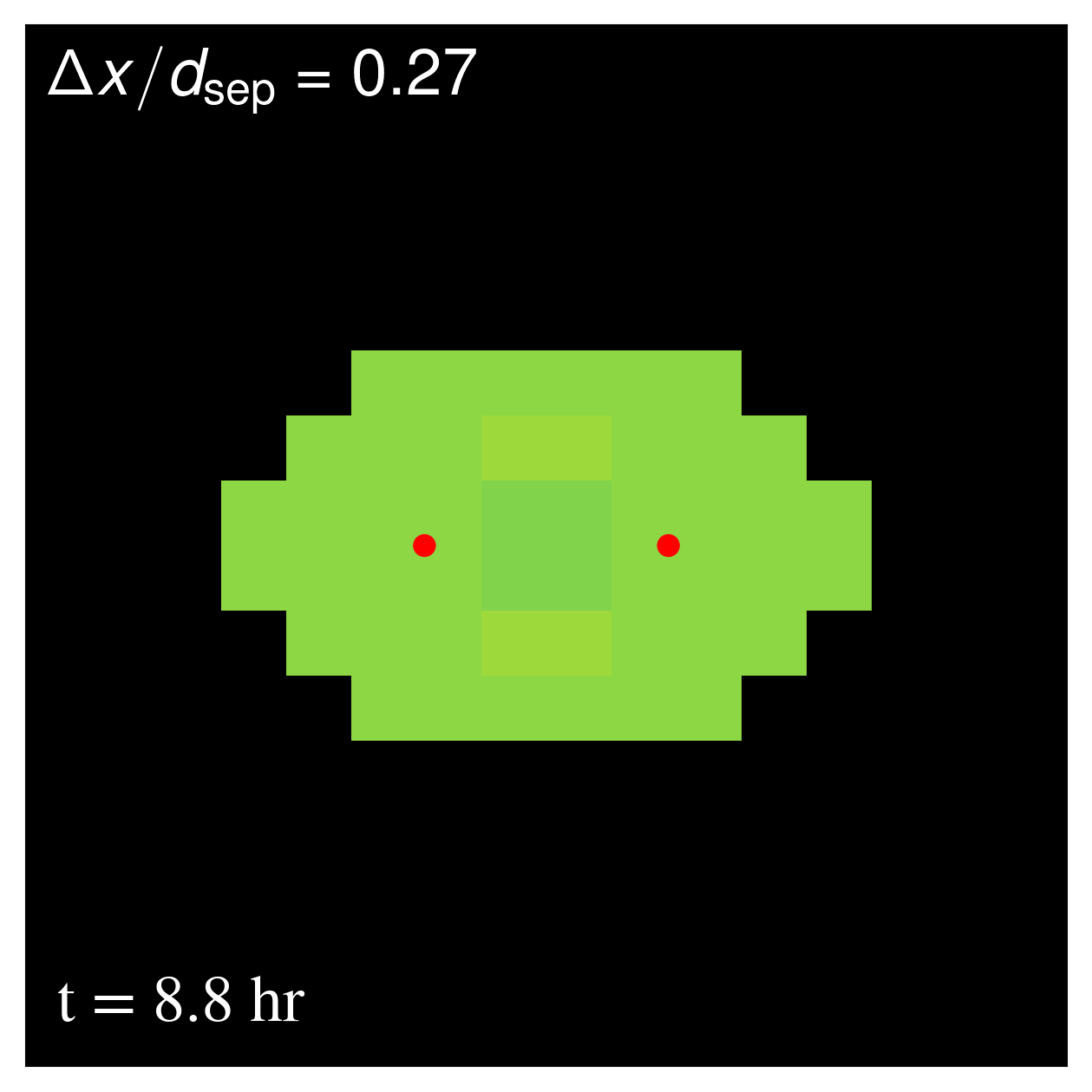}\includegraphics[scale=0.34]{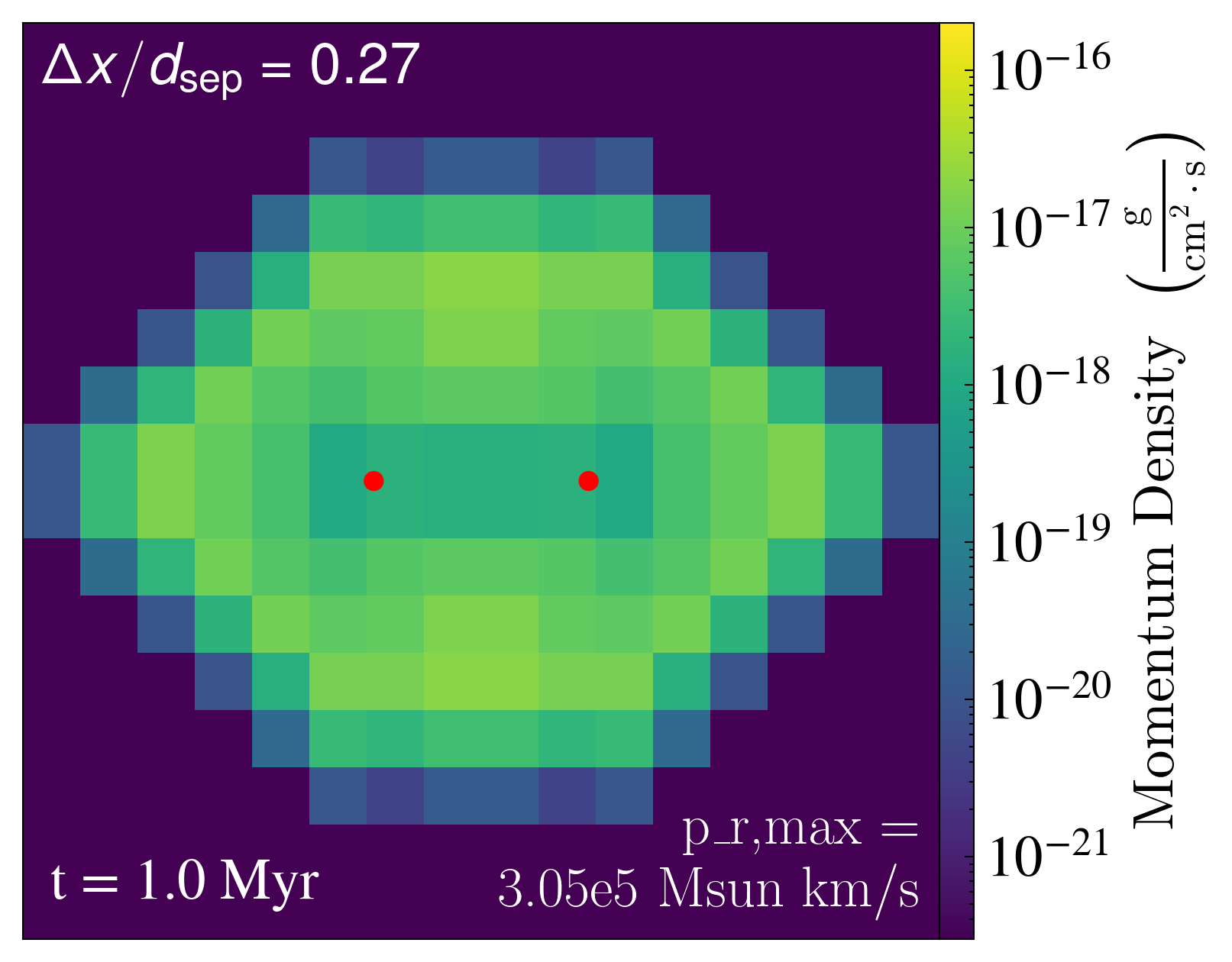}\\
    \includegraphics[scale=0.34]{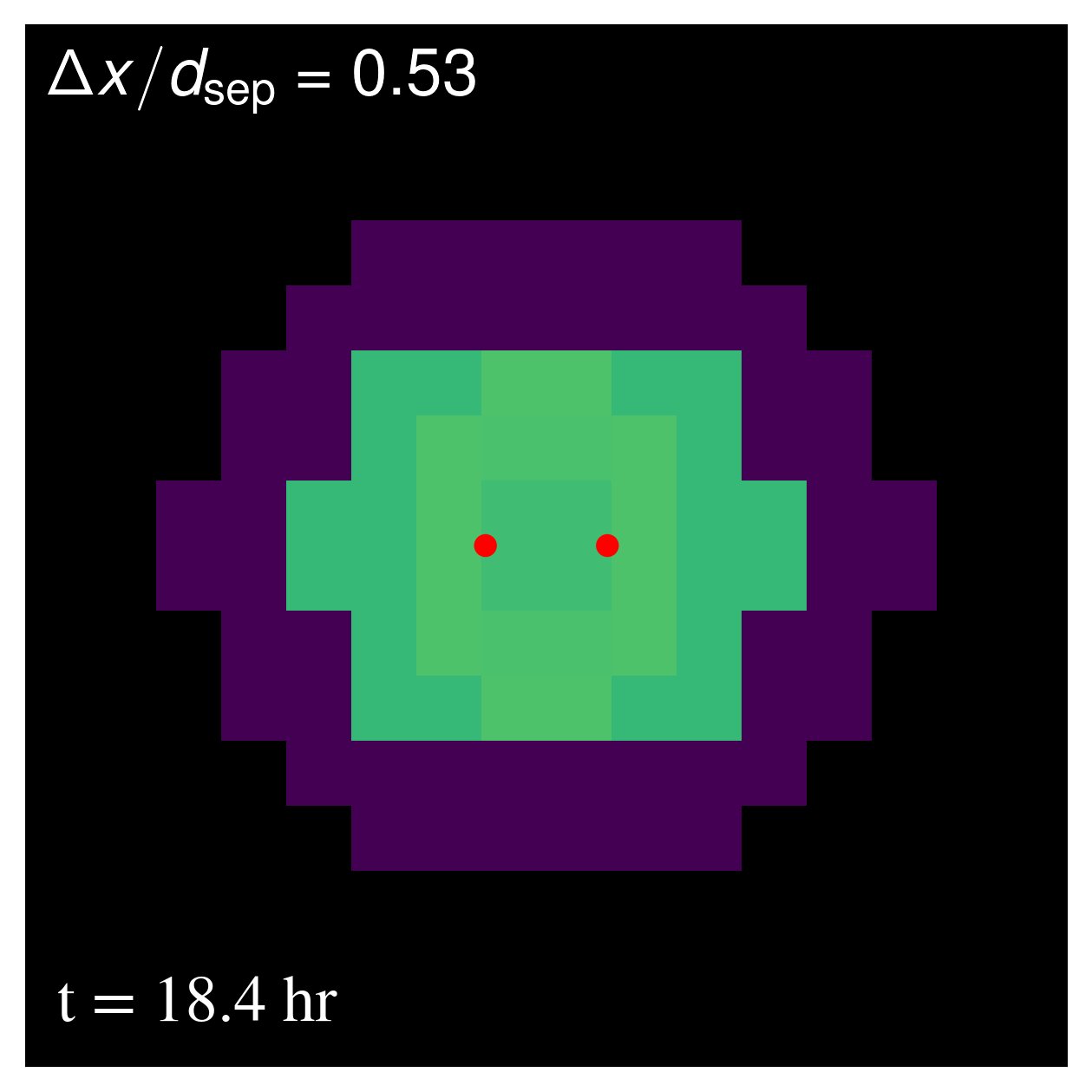}\includegraphics[scale=0.34]{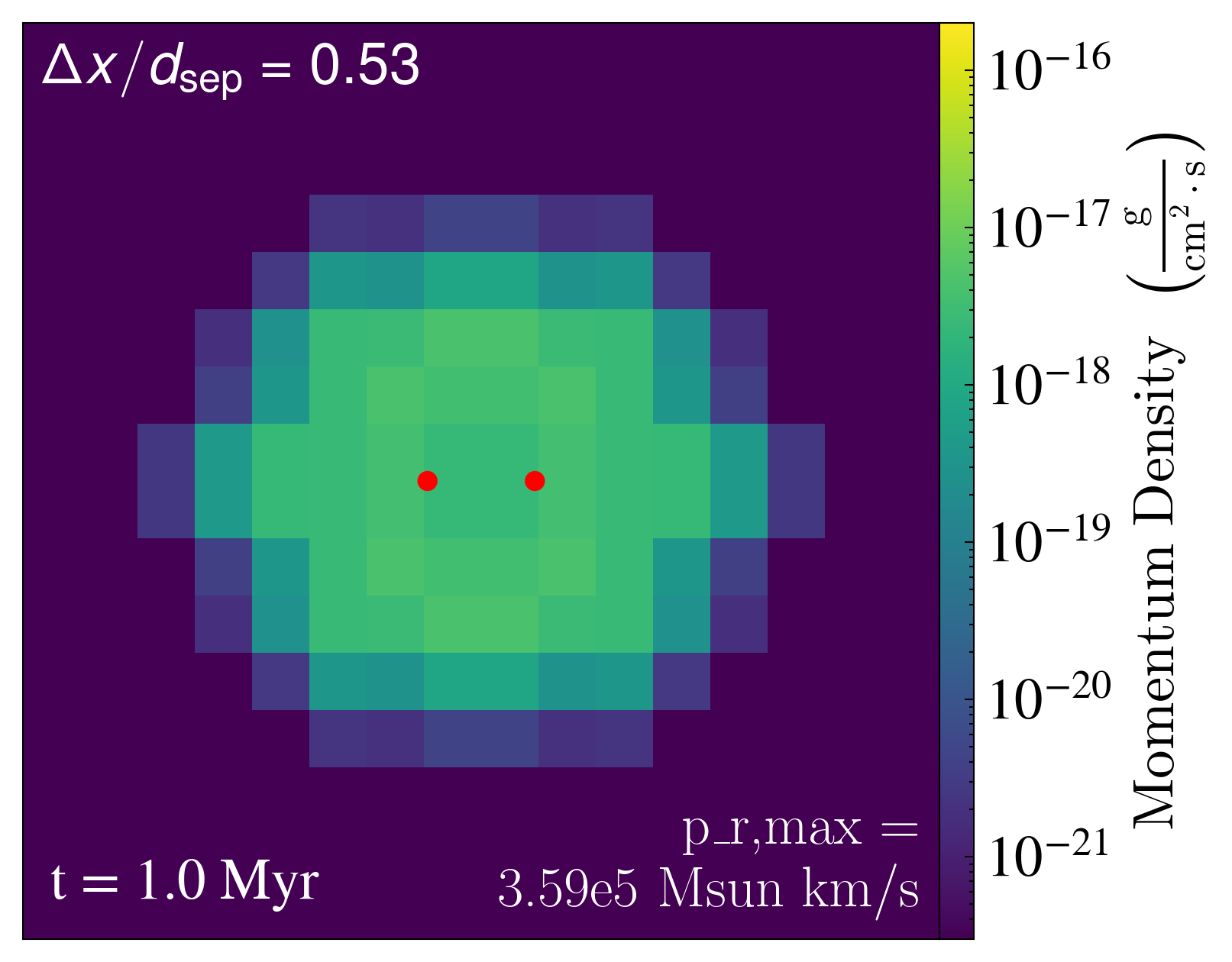}
    \caption{Same as \autoref{fig:sn22} but for a setup where $R_{\rm shell} / d_{\rm sep} = 0.3$ (corresponding to an ambient density $n_\mathrm{H,amb} = 10$ cm$^{-3}$). The full simulation box is 256 pc in length, but for improved visibility the first three rows are cropped to a 128 pc region. In this configuration, shell formation occurs {\it before} SNRs begin to interact. Our numerical method demonstrates convergent terminal radial momentum across spatial resolutions, with discrepancies within $20\%$.
    }
    \label{fig:sn21}
\end{figure}

\begingroup % Optional: affects only this particular table
    \renewcommand{\arraystretch}{1.5} % Default value: 1
\begin{table*}
    \centering
    \begin{tabular}{cccccccc}
        \hline\hline
         $n_{\rm H, amb}$& $r_{\rm sf}$& $\Delta x$ & $\Delta x / d_{\rm sep}$ & $2 r_{K} / d_{\rm sep}$ & $r_K / r_{\rm sf} $&$p_{\rm r,max}$ &$\epsilon_p$\\
 $[{\rm cm}^{-3}]$& [pc]& [pc]& & & &$[{\rm M}_{\odot}~{\rm km}~{\rm s}^{-1}]$ &$[\%]$\\
         \hline
 0.1& 59& 2& 0.067& 0.40& 0.1&$6.92 \times 10^{5}$ &-\\
 0.1& 59& 4& 0.13& 0.80& 0.2&$6.96 \times 10^{5}$ &0.6\\
 0.1& 59& 8& 0.27& 1.60& 0.4&$6.73 \times 10^{5}$ &$-2.7$\\
 0.1& 59& 16& 0.53& 3.20& 0.8&$6.69 \times 10^{5}$ &$-3.3$\\
 10& 8.6& 2& 0.067& 0.40& 0.7&$3.00 \times 10^{5}$ &-\\
 10& 8.6& 4& 0.13& 0.80& 1.4&$3.42 \times 10^{5}$ &14\\
 10& 8.6& 8& 0.27& 1.60& 2.8&$3.05 \times 10^{5}$ &1.7\\
 10& 8.6& 16& 0.53& 3.20& 5.6&$3.59 \times 10^{5}$ &20\\
    \hline\hline
    \end{tabular} 
    \caption{Parameters and results of the numerical convergence tests for the evolution of two SNRs separated by $d_{\rm dep} = 30$ pc. Symbols: $n_{\rm H, amb}$ -- ambient gas number density;  $r_{\rm sf}$ -- shell-formation radius, $\Delta x$ -- spatial resolution; $p_{\rm r,max}$ -- total terminal radial momentum; $\epsilon_p$ -- error relative to the value at the best resolution for a given $n_{\rm H, amb}$. In all tests, the SN separation is 30 pc. The strength of SN feedback, characterized by $p_{\rm r,max}$, matches the high-resolution case to within 20 \% for all cases and resolutions.}
    \label{tab:sn2}
\end{table*}
\endgroup

One advantage of our algorithm with a limiter is its robust treatment of multiple SNe depositing into a single cell. To test numerical convergence in overlapping SNR regions, we simulate the evolution of two SNe exploding simultaneously at $t=0$. We consider two ambient densities, $0.1~{\rm cm}^{-3}$ and $10~{\rm cm}^{-3}$, corresponding to shell-formation radii of 59 pc and 8.6 pc, respectively. We place the exploding particles symmetrically about the domain centre, separated by $d_{\rm sep} = 30$ pc, which lies between the two values of $r_{\rm sf}$: for the two lower and higher initial densities, $r_\mathrm{sf}/d_\mathrm{sep} \approx 2.0$ and $0.3$, respectively. Consequently, for the higher ambient density the SNRs should form shells before they begin to interact, while for the lower ambient density interaction will begin during the Sedov-Taylor phase.

We simulate each of our two configurations at spatial resolutions $\Delta x = 2, 4, 8$ and $16$ pc in a domain of size 256 pc, again with no adaptivity. Given the particle separation of 30 pc and our fiducial kernel size of $r_K = 3 \Delta x$, the deposition kernels of the two particles are well-separated in the higher-resolution runs, but overlap in the lower-resolution runs. Comparing the low- and high-resolution cases therefore allows us to test the performance of our scheme for handling multiple SNe in cases where the deposition regions for those SNe overlap, in both the well-resolved, $r_\mathrm{sf}/r_K \gg 1$, and poorly-resolved, $r_\mathrm{sf}/r_K \lesssim 1$, regimes. We summarise the full set of configurations we have tested in \autoref{tab:sn2}. 

\autoref{fig:sn22} and \autoref{fig:sn21} illustrate the momentum-density evolution across resolutions and initial densities. Consistent with our expectations, we see shell formation prior to interaction between the two SNRs at higher density and when the resolution is sufficient for the deposition kernels not to overlap, while for lower density the two SNRs merge before forming shells.

Because there is no empirical $p_{\rm r,snr}-n_{\rm H, amb}$ relation for the dual-SNR problem, we take the 2 pc-resolution run as the reference solution for each ambient density. We measure the terminal radial momentum in that simulation, and report the fractional deviation from that value in \autoref{tab:sn2} as our error estimate. For $n_{\rm H, amb} = 10~{\rm cm}^{-3}$, the reference solution lies in the regime where the ST phase is resolved and the initial injection regions do not overlap. For $n_{\rm H, amb} = 0.1~{\rm cm}^{-3}$, the shells of the two SNRs overlap, but the reference solution still resolves the ST phase and avoids overlap at the moment of injection. Across all resolutions and both ambient densities, the terminal radial momentum $p_{\rm r,max}$ matches the high-resolution result to within $20\%$, demonstrating the effectiveness of the limiter.

\subsection{Parallel performance}

\begin{figure}
  \centering
  \includegraphics[width=\columnwidth]{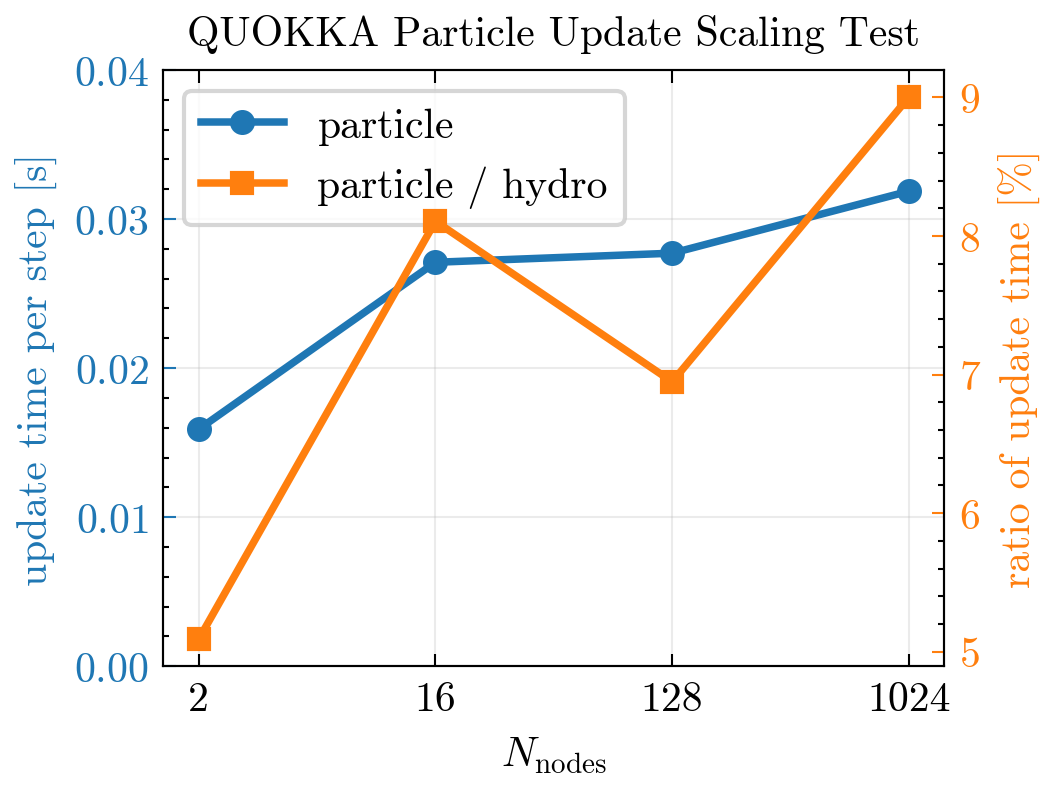}
  \caption{\label{fig:scaling} Weak-scaling test of the particle update in \quokka{} for a Milky Way--like disk galaxy simulation. The left $y$ axis shows the wall-clock time for the particle update per coarse step, and the right $y$ axis shows this time expressed as a fraction of the hydrodynamic update time, both as a function of the number of Frontier nodes. We find 50\% weak scaling efficiency from 2 to 1024 nodes.
  }
\end{figure}

\begingroup % Optional: affects only this particular table
    \renewcommand{\arraystretch}{1.5} % Default value: 1
\begin{table}
  \centering
    \begin{tabular}{rrrr}
        \hline\hline
        Level & $\Delta x$ [pc] & $N_\mathrm{cell}^{1/3}$ & $V^{1/3}$ [kpc] \\
        \hline
        0 & 1171.9 & 1024 & 1200 \\
        1 & 585.9 & 1024 & 600 \\
        2 & 293.0 & 1024 & 300 \\
        3 & 146.5 & 1024 & 150 \\
        4 & 73.2 & 1024 & 75 \\
        5 & 36.6 & 1342 & 49 \\
        6 & 18.3 & 1886 & 35 \\
        7 & 9.2 & 2954 & 27 \\
        8 & 4.6 & 4721 & 22 \\
        \hline\hline
    \end{tabular}
    \caption{\label{tab:scaling} Parameters of the weak-scaling benchmark at the highest node count of 1024. For each AMR level we list the cell size $\Delta x$, the cube root of the total cell count $N_\mathrm{cell}^{1/3}$, and the cube root of the simulated volume $V^{1/3}$. The numbers given are for the highest node count of 1024; for smaller node counts the volume $V$ remains unchanged, but the cell size $\Delta x$ increases in factor of 2 steps, and the cell and particle counts decrease in factor of $2^3$ steps.
    The total number of particles in the simulation is $1.088 \times 10^9 = 1029^3$, and about half of them are star particles at the finest level.}    
\end{table}
\endgroup

We evaluate the parallel performance of the \quokka{} particle module using weak-scaling benchmarks on the Frontier supercomputer at Oak Ridge National Laboratory. Each node of this system has 8 AMD MI-250X logical GPUs. In a weak-scaling experiment the total problem size is increased in proportion to the number of compute nodes so that the workload per node remains approximately fixed; ideal weak scaling therefore corresponds to constant wall-clock time per step as the node count increases.

We consider an isolated Milky Way analogue initial condition from the AGORA project \citep{Kim2016a}, a standardised equilibrium disk galaxy model comprising a gas disk, a stellar disk and bulge, and a dark matter halo with masses $M_\mathrm{gas} \approx 9 \times 10^{9}$ M$_\odot$, $M_\mathrm{*,disk+bulge} \approx 4 \times 10^{10}$ M$_\odot$, and $M_\mathrm{halo} \approx 10^{12}$ M$_\odot$. The stellar and dark matter components are represented by live collisionless particles that interact gravitationally with the gas and with one another, and are initialised in equilibrium using a numerical solution of the Jeans equations.

This benchmark uses the full physics stack adopted in our production galaxy simulations: piecewise parabolic method hydrodynamics with an RK2 update that is second-order accurate in time and space, multigrid self-gravity, radiative cooling using a GPU Runge-Kutta ODE integrator with the \textsc{Grackle} tabulated cooling and heating rates \citep{Smith2017} with temperature evolution integrated to a relative tolerance of $10^{-6}$, and the star-formation and supernova feedback prescriptions described in \autoref{sec:impl}. We enable AMR and construct a sequence of problem sizes that preserves the refinement pattern while increasing the total problem size with node count.

For this test all runs use the same nine-level AMR hierarchy (levels 0--8). We generate the 2-, 16-, 256-, and 1024-node cases (corresponding to 8 to 8192 GPUs) by resampling a checkpoint so that the refinement geometry is preserved but the resolution is increased uniformly at all AMR levels. Across this sequence, the linear resolution increases by factors of 2, 4, and 8 relative to the two-node case (and the number of cells correspondingly increases by factors of $2^3$, $4^3$, and $8^3$), yielding finest-level cell sizes of $\Delta x = 36.6, \ 18.3, \ 9.2$, and $4.6$ pc, respectively. We similarly scale the particle load by increasing the particle number and decreasing the particle masses by factors of $2^3$, $4^3$, and $8^3$ across the three resolution increases. We report counts for the numbers of cells on each AMR level for the 1024-node case in \autoref{tab:scaling}. The total number of particles in the simulation domain for this node count is $1.088 \times 10^9$; about half of these are star particles located in the disk that overlap with cells refined to the maximum level, and the other half are dark matter particles distributed among all levels. 

To test the performance of the particle code, we measure the wall-clock time of the particle update per step, which we obtain from an inline profiler in our code that isolates the particles routines, by averaging over 100 steps with I/O disabled. The results are shown in \autoref{fig:scaling}; the blue line shows the absolute update time per step, while the orange line shows the ratio of the time spent updating particles to the time spent updating hydrodynamics. Since in this test the numbers of particles and cells per GPU are fixed, perfect weak scaling corresponds to a horizontal line. We measure a weak scaling efficiency of $50\%$ from 2 to 1024 nodes for the particle code, but even at the largest node count the particle update requires less than $10\%$ of the hydrodynamic update time. This good scaling is achieved because the star-formation and feedback modules introduced in this work execute natively on GPUs without GPU--CPU transfers of memory, and because the particle module avoids all-to-all communication by restricting communication to the same nearest-neighbour ghost-zone exchanges required by the hydrodynamic update.

\section{Summary}
\label{sec:summary}

We present a new algorithm, particle‑mesh‑particle, to provide a robust and efficient solution to the challenges inherent in particle–mesh and particle-particle interactions on modern GPU architectures. By avoiding expensive neighbour searches through the use of a deposition-buffer mesh with ghost zones and using efficient space-filling communication of these zones, our method achieves both high numerical accuracy and excellent performance. We provide implementations of two types of particle-mesh interaction -- sink particles and SN feedback particles -- built around this algorithm, and show that these implementations pass multiple tests, including accretion from a singular isothermal sphere, Bondi and Bondi-Hoyle flows, and expansion of SNRs driven by single and multiple SNe at a range of resolutions. To our knowledge, this approach enables, for the first time, simulations of star formation and stellar feedback on massively parallel, GPU-based architectures. The algorithm is also readily extensible to other feedback processes, such as stellar winds and protostellar outflows, and more generally to any linear or nonlinear particle-particle or particle-mesh interactions.

Our particle‑mesh‑particle strategy represents a paradigm shift by rethinking how particle contributions are aggregated and updated on GPUs. It requires only one extra communication per hydrodynamic step and reuses the communication pattern of the hydrodynamic update, which is already well-optimized for GPUs. We demonstrate the computational efficiency of this approach through weak-scaling benchmarks on the Frontier supercomputer using full-physics galaxy simulations with adaptive mesh refinement, self-gravity, radiative cooling, star formation, and supernova feedback. We find that the particle module demonstrates $\approx 50\%$ weak scaling from 2 to 1024 nodes (8192 GPUs). This performance demonstrates that our method scales comparably to pure hydrodynamics and can run efficiently on thousands of GPUs.

Despite these promising results, several limitations remain. First, the interaction distance is constrained by the number of hydrodynamic ghost cells. For complex stellar-feedback prescriptions that combine multiple feedback mechanisms (e.g., sink accretion, stellar wind, and outflows), additional ghost cells may be required to maintain accuracy. While increasing the number of ghost cells beyond what is required for the hydrodynamic solver can improve the accuracy of particle feedback, it incurs additional memory and compute costs.
Second, although we have minimized inter-GPU communications to a single ghost-zone synchronization per hydrodynamic step, this constitutes an additional synchronization beyond that needed for the hydrodynamic update. This ghost zone update must be applied on levels that host particles and across all domains, regardless of the number of particles, and there is no obvious way to merge this synchronization with that of the hydrodynamic solver.

Future work will focus on extending \quokka{}'s particle module to include additional feedback processes such as stellar winds and protostellar outflows. GPU-optimized methods for particle-particle-particle-mesh $N$-body calculations will also be explored.

\section*{Acknowledgements}

CCH, AV, and MRK acknowledge support from the Australian Research Council through Laureate Fellowship FL220100020. This research was undertaken with the assistance of resources and services from the National Computational Infrastructure (NCI) and the Pawsey Supercomputing Centre, which are supported by the Australian Government, through award jh2. PSL acknowledges the supports by the NSFC through grant No. 1241101426 and the National Key R\&D Program of China (No. 2022YFA1603101). The \quokka{} code is based on the AMReX module \citep{AMReX_JOSS}.

\section*{Data Availability}

The \quokka{} code, including the source code for all tests presented in this paper, is available from \url{https://github.com/quokka-astro/quokka} under an open-source license.

\bibliographystyle{mnras}
\bibliography{ref}

\appendix

\section{Bitwise-reproducibility}
\label{app:reproducibility}

We define software as bitwise-reproducible if it has the property that running it multiple times using the same inputs (including seeds for any random number generators) on the same hardware (i.e., using the same number and configuration of GPUs) is guaranteed to produce outputs that are bit-by-bit identical. This property is desirable for numerical software because it greatly aids in debugging and correctness testing -- for example, if software is reproducible then we can verify that code changes that should not affect the outcomes of particular tests indeed do not affect them by performing bitwise differences between results produced before and after the change. However, bitwise reproducibility is very difficult to guarantee for GPU-based software, and strategies for maintaining it are an active area of research in the computer science community \citep[e.g.][]{Defour15a, Ahrens20a}.

Bitwise-reproducibility is challenging on GPU due to the interaction between the unpredictable order of execution of different GPU threads and the non-associativity of floating point arithmetic. The latter property means that, when we sum a list of floating point numbers, the sum is not guaranteed to be bitwise-invariant under a re-ordering of the list, while the former means that if the sum is being computed by multiple GPU threads, or if insertion of numbers into the list is carried out by multiple threads, we cannot guarantee that the list will be summed in the same order every time the program is run. This means that a naive program that runs reproducibility on a single CPU thread will not be reproducible when run on GPU if it involves any summations of lists of floating point numbers or analogous operations. The first two steps of the algorithm described in \autoref{ssec:alg_steps} both involve summations that, if not carried out carefully, would lead to non-reproducibility. We limit this problem as follows.

\subsection{Particle to buffer mesh}

The first step in our algorithm is to sum the contributions $\Delta\vecU_{ijk}^s$ for all particles $s$ to the buffer mesh, which is exactly the type of list summation operation for which a naive implementation will not be reproducible on GPU. To improve reproducibility, we modify our deposition algorithm as follows:
\begin{enumerate}
    \item In addition to creating a buffer to hold the amount of conserved quantity we are depositing in each cell $\Delta \vecU_{ijk}^\mathrm{buf}$, we also create a second temporary buffer $\vecE_{ijk}$ covering the same cells, which we will use to accumulate an upper bound on the error due to non-associativity.
    \item When we update $\Delta \vecU_{ijk}^\mathrm{buf}$ by depositing the contribution from a given particle to it, we also update the value of $\vecE$ for the corresponding cell by carrying out the operations
    \begin{eqnarray*}
        \Delta\vecU_{ijk}^\mathrm{buf} & := & \Delta\vecU_{ijk}^\mathrm{buf} + \Delta\vecU_{ijk}^s \\
        \vecE_{ijk} & := & \vecE_{ijk} + \varepsilon \left|\Delta\vecU_{ijk}^\mathrm{buf} \right|,
    \end{eqnarray*}
    where $\varepsilon$ is the machine precision. The amount we are adding to $\vecE$ represents an upper limit on the amount by which the sum could be changed by re-ordering the additions.
    \item After we have completed summing over all particles $s$, we compute an upper bound on the relative error in each cell as $\mathbf{r}_{ijk} = \vecE_{ijk} / |\Delta\vecU_{ijk}^\mathrm{buf}|$.
    \item We then round $\Delta\vecU_{ijk}^\mathrm{buf}$ in every cell by setting the least significant $N_\mathrm{bit}$ bits of the floating point mantissa to zero, with $N_\mathrm{bit} = \log_2 \mathbf{r}_{ijk} + N_\mathrm{mantissa} + M$, where $N_\mathrm{mantissa}=52$ is the number of bits in the mantissa and the extra $+M$ is a safety margin.
\end{enumerate}

The choice of $M$ requires some thought, and involves a trade-off between reproducibility and accuracy. Naively one might expect that choosing any $M \gtrsim 1$ would be sufficient to ensure reproducibility, since in this case the amount by which the results could differ between successive runs the program as a result of re-ordering of summations is much smaller than the amount by which we are rounding. However, rounding involves mapping all the bit values within a certain interval to a single value, and there is always a chance that, when the code is run twice, the result from the first run will be so close to the boundary between rounding intervals that the small amount by which the results differ in the second run pushes the sum into a different rounding interval. The probability of this happening is roughly the amount by which the results can change between runs divided by the size of the rounding interval, and thus the probability is $p\approx 2^{-M}$. We can therefore choose $M$ large enough that, for a particular application, it is very unlikely that a non-reproducible result will occur within a specified run duration. The value of $M$ required for this purpose depends on the typical number of cells in which there are contributions from multiple particles and on the number of time steps for which the simulation is expected to run. For example the choice $M = 28$, which for cases where at most a few particles contribute to each cell is roughly equivalent to rounding the results from double precision to single precision, is sufficient to make a non-reproducible result improbable as long as the number of cell-advances during which multiple particles deposit to the same cell is $\lesssim 10^9$. In future work we will also explore the option of using superaccumulators, as suggested by \citet{Defour15a} and \citet{Collange15a}, to guarantee full bitwise-reproducibility for an arbitrary number of steps.

\subsection{Summation of the buffer mesh}

The second location in our algorithm where potential non-reproducibility arises is in the summation of the buffer mesh, where we add the buffer meshes that contain the contributions from all particles that can influence a given cell. As shown in \autoref{fig:pmp}, for most buffer cells there are at most two contributions, and thus there is no problem with associativity, but for cells that lie near the corner (in 2D) or edge (in 3D) of a cubical domain, there may be more than two buffers that contribute to that cell, in which case the potential for non-reproducibility appears.

To handle this problem we flag cells in the overlapping region over which the sum is taking place to which more than two different buffers contribute. We sum the majority cells with only two contributors immediately, and for the remaining cells we accumulate the contributions from all buffers in memory, sorting them based on the $(x,y,z)$ position of the contributing grid, and then sum only once all the contributions have been transferred. Since the ordering of the summation is then uniquely determined and guaranteed to be the same every time the program is run, the result is bitwise-reproducible.

\end{document}